\documentclass[referee]{aa} 
\usepackage{graphicx}
\usepackage{txfonts}
\usepackage{natbib}
\usepackage{placeins}
\newcommand{\solm}{M$_{\odot}$}

\newcommand\arcdeg{\mbox{$^\circ$}}%
\begin{document}
%
\title{The Flare Activity of Sgr~A*}

   \subtitle{New Coordinated mm to X-Ray Observations}

   \author{A. Eckart\inst{1}
	  \and
          F. K. Baganoff\inst{2}
          \and 
          ~R. Sch\"odel\inst{1}
          \and 
          M. Morris\inst{3}
          \and
          R. Genzel\inst{4,5}
          \and
          G.C. Bower \inst{6}
          \and
          D. Marrone \inst{7}
          \and
          J.M. Moran \inst{7}
          \and
          ~T. Viehmann\inst{1}
          \and
          M.W. Bautz\inst{2}
          \and
          W.N. Brandt\inst{8}
          \and
          G.P. Garmire\inst{8}
          \and
          T. Ott\inst{4}
          \and
          S. Trippe\inst{4}
          \and
          G.R. Ricker\inst{2}
          \and
          C. Straubmeier\inst{1}
          \and
          D.A. Roberts  \inst{9}
          \and
          F. Yusef-Zadeh  \inst{9}
          \and
          J.H. Zhao \inst{7}
          \and
          R. Rao \inst{7}
          }

   \offprints{A. Eckart}

   \institute{ I.Physikalisches Institut, Universit\"at zu K\"oln,
              Z\"ulpicher Str.77, 50937 K\"oln, Germany\\
              \email{eckart@ph1.uni-koeln.de}
         \and
             Center for Space Research, Massachusetts Institute of
              Technology, Cambridge, MA~02139-4307, USA\\
             \email{fkb@space.mit.edu}
         \and
             Department of Physics and Astronomy, University of 
             California Los Angeles, Los Angeles, CA~90095-1562, USA\\
             \email{morris@astro.ucla.edu}
         \and
             Max Planck Institut f\"ur extraterrestrische Physik,
              Giessenbachstra{\ss}e, 85748 Garching, Germany
         \and
             Department of Astronomy and Radio Astronomy Laboratory,
             University of California at Berkeley, Le Conte Hall, 
             Berkeley, CA~94720, USA
         \and
             Department of Astronomy and Radio Astronomy Laboratory,
             University of California at Berkeley, Campbell Hall, 
             Berkeley, CA~94720, USA\\
             \email{gbower@astro.berkeley.edu}
         \and
             Harvard-Smithsonian
             Center for Astrophysics, Cambridge MA 02138, USA\\
             \email{dmarrone.cfa.harvard.edu}
         \and
             Department of Astronomy and Astrophysics, Pennsylvania
              State University, University Park, PA~16802-6305, USA
         \and
             Department of Physics and Astronomy, 
             Northwestern University, Evanston, IL 60208
             }

   \date{Received ; Accepted }

   \date{}

\abstract{
We report new simultaneous near-infrared/sub-millimeter/X-ray
observations of the SgrA* counterpart associated with the
massive 3--4$\times$10$^6$\solm ~black hole at the Galactic Center.  
}{
The main aim is to investigate the physical processes 
responsible for the variable emission from SgrA*.
}{
The observations have been carried out using the NACO adaptive
optics (AO) instrument at the European Southern Observatory's Very Large
Telescope\footnote{Based on observations at the Very Large Telescope
(VLT) of the European Southern Observatory (ESO) on Paranal in Chile;
Program: 271.B-5019(A).} and the ACIS-I instrument aboard the
\emph{Chandra X-ray Observatory} as well as the Submillimeter Array
SMA\footnote{The Submillimeter
Array is a joint project between the Smithsonian Astrophysical
Observatory and the Academia Sinica Institute of Astronomy and
Astrophysics, and is funded by the Smithsonian Institution and the
Academia Sinica.} on Mauna Kea, Hawaii,
and the Very Large Array\footnote{The VLA is operated by the 
National Radio Astronomy Observatory 
which is a facility of the National Science Foundation operated 
under cooperative agreement by Associated Universities, Inc.}
in New Mexico.  
}{
We detected one moderately 
bright flare event in the X-ray domain and 5 events at 
infrared wavelengths. 
The X-ray flare had 
an excess 2 - 8 keV luminosity of about
33$\times$10$^{33}$~erg/s. 
The duration of this flare was completely covered in the infrared and
it was detected as a simultaneous NIR event with a time lag of 
$\le$10~minutes.  
For 4 flares simultaneous infrared/X-ray observations are available. 
All simultaneously covered flares, combined with the flare covered in
2003, indicate that the time-lag between the NIR and X-ray flare emission is
very small and in agreement with a synchronous evolution.
There are no simultaneous flare detections between the NIR/X-ray data and
the VLA and SMA data. The excess flux densities detected in the 
radio and sub-millimeter domain may be linked with the flare activity
observed at shorter wavelengths.
}{
We find that the flaring state can be explained
with a synchrotron self-Compton (SSC) model involving up-scattered
sub-millimeter photons from a compact source component.
This model allows for NIR flux density contributions 
from both the synchrotron and SSC mechanisms.
Indications for 
an exponential cutoff of the NIR/MIR synchrotron spectrum 
allow for a straight forward explanation of 
the variable and red spectral indices of NIR flares.
}


   \titlerunning{Simultaneous NIR/X-ray Flare from Sgr~A*}
   \authorrunning{Eckart, Baganoff, Morris, Sch\"odel  et al.}  
   \maketitle
%

\section{Introduction}
\label{section:Introduction}

Over the last decades, evidence has been accumulating that most quiet
galaxies harbor a massive black hole (MBH) at their centers.
Especially in the case of the center of our Galaxy, progress could be
made through the investigation of the stellar dynamics (Eckart \&
Genzel 1996, Genzel et al. 1997, 2000, Ghez et al. 1998, 2000, 2003a,
2003b, 2005, Eckart et al. 2002, Sch\"odel et al. 2002, 2003, Eisenhauer 2003, 2005).  
At a distance of only $\sim$8 kpc from the sun (Reid 1993, Eisenhauer et
al. 2003, 2005), the Galactic Center allows for detailed observations 
of stars at distances much
less than 1~pc from the central black hole candidate, the compact radio 
source Sgr~A*.  Additional compelling evidence for a massive black hole at the
position of Sgr~A* is provided by the observation of variable emission 
from that position both in the
X-ray and recently in the near-infrared wavelength domain (Baganoff et
al. 2001, 2002, 2003, Eckart et al. 2003, 2004, Porquet et al. 2003, Goldwurm et
al. 2003, Genzel et al. 2003, Ghez et al. 2004a, and Eisenhauer et al. 2005).  
Throughout the paper
we will use the term 'interim-quiescent' (or IQ) for the 
phases of low-level, and especially in the NIR domain - possibly 
continously variable flux density 
at any given observational epoch. 
This state may represent flux density variations 
on longer time scales (days to years). This is especially
true for the NIR source (Genzel et al. 2003, Ghez et
al. 2004a, Eckart et al. 2004).

Simultaneous observations of SgrA* across different wavelength regimes
are of high value, since they provide information on the emission
mechanisms responsible for the radiation from the immediate vicinity
of the central black hole.  The first observation of SgrA* detecting
an X-ray flare simultaneously in the near-infrared was presented by 
Eckart et al. (2004). They detected a weak 
6$\times$10$^{33}$~erg/s X-ray flare and covered its decaying flank 
simultaneously in the NIR. 

Variability at radio through submillimeter wavelengths has been studied
extensively, showing that variations occur on time scales from hours to
years (Wright \& Backer 1994,
Bower et al. 2002, Herrnstein et al. 2004, Zhao et al. 2003).  Some
of the variability may be due to interstellar scintillation.  The
connection to variability at NIR and X-ray wavelengths has not been
clearly elucidated.  Zhao et al. (2004) showed a probable link between the
brightest X-ray flare ever observed and flux density at 
0.7, 1.3, and 2 cm
wavelength on a timescale of $<1$ day (see also Mauerhan et al. 2005).

In section 2 of the present paper we report on
new simultaneous NIR/X-ray observations using
\emph{Chandra} and the adaptive optics instrument NACO at the VLT UT4.  
The new 8.6$\mu$m and 19.5$\mu$m
observations of the central region were obtained during
the commissioning of the ESO MIR VISIR camera. 
We also describe the new SMA and VLA data of SgrA*.
These data give additional information
on the flux density limit of SgrA* at millimeter and sub-millimeter wavelengths.
In section 3 we discuss the NIR to X-ray variability of SgrA*,
followed by a discussion of its MIR/NIR spectrum in section 4.
In sections 5 and 6 we discuss the flux densities, spectral indices and flares 
observed in the NIR and X-ray domain.
The physical interpretation in section 7 is then followed by the summary and discussion 
in the final section 8.


\section{Observations and Data Reduction}
\label{section:Observations}

\begin{figure}
\centering
\includegraphics[width=15cm]{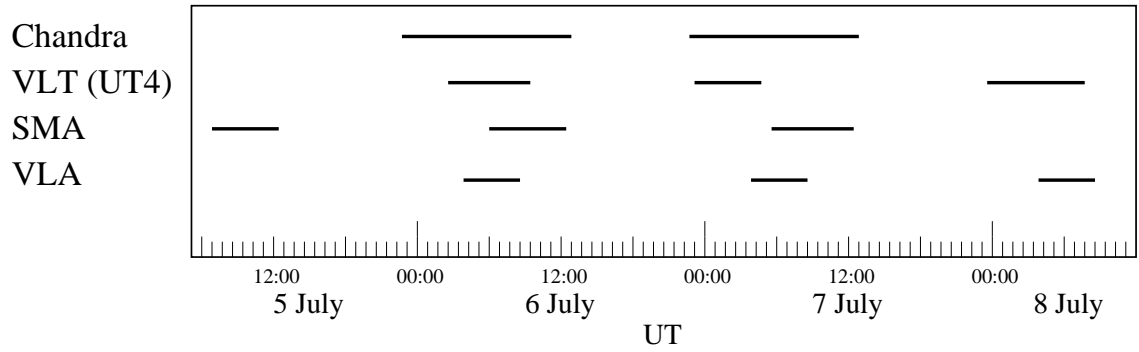}
\caption{Schematic view of the observing schedule. 
The exact times are listed in Tab.~\ref{log}.
\label{Fig:schedule}
}
\end{figure}

Sgr~A* was observed from the radio millimeter to the X-ray wavelength
domain. 
Fig.~\ref{Fig:schedule} shows the schematic observing schedule and
Tab.~\ref{log} lists the individual observing sessions.
In the following we describe the data acquisition and reduction 
for the individual telescopes.

\subsection{The NACO NIR Adaptive Optics Observations}
\label{section:NACO}

Near-infrared (NIR) observations of the Galactic Center (GC) were
carried out with the NIR camera CONICA and the adaptive optics (AO)
module NAOS (briefly ``NACO'') at the ESO VLT unit telescope~4 on
Paranal, Chile, during the nights between 05 July and 08 July 2004. In all
observations, the infrared wavefront sensor of NAOS was used to lock
the AO loop on the NIR bright (K-band magnitude $\sim$6.5) supergiant
IRS~7, located about $5.6''$ north of Sgr~A*.  The start and stop
times of the NIR observations are listed in Table~\ref{log}. Details
on integration times and approximate seeing during the
observations are listed in Table~\ref{NIRObs}. 
As can be seen, the
atmospheric conditions (and consequently the AO correction) were
fairly variable during some of the observation blocks.

Observations of a dark cloud -- a region practically empty of stars --
a few arcminutes to the north-west of Sgr~A* were interspersed with
the observations at 1.7\,$\mu$m and 2.2\,$\mu$m in order to obtain sky
measurements. All observations were dithered in order to cover a
larger area of the GC by mosaic imaging. In case of the observations
at 3.8\,$\mu$m, the sky background was extracted from the median of
stacks of dithered exposures.
Here the procedure is different from that at 1.7\,$\mu$m and 2.2\,$\mu$m
since thermal emission from dust as well as a brighter and variable
sky has to be taken into account.
All exposures were sky subtracted, flat-fielded, and corrected for
dead or bad pixels. 
In order to enhance the signal-to-noise ratio
of the imaging data, we therefore created median images comprising 9
single exposures each. Subsequently, PSFs were extracted from these
images with \emph{StarFinder}
(Diolaiti et al. 2000).
The images were deconvolved with the
Lucy-Richardson (LR) and linear Wiener filter (LIN) algorithms. Beam
restoration was carried out
with a Gaussian beam of FWHM corresponding to the respective
wavelength. Therefore the final resolution at 1.7, 2.2, and 3.8 \,$\mu$m
is 46, 60, and 104~milli-arcseconds, respectively.

The flux densities of the sources were measured by aperture photometry with
circular apertures of 52~mas radius and corrected
for extinction, using $A_{H} = 4.3$,  $A_{K} = 2.8$, and  $A_{L'} = 1.8$.
Calibration of the photometry and
astrometry was done with the help of the known fluxes and positions of
9 sources within $1.6"$ of Sgr~A*. Uncertainties were obtained by
comparing the results of the photometry on the LR and LIN deconvolved
images. The background flux was obtained by averaging the measurements
at five random locations in a field about $0.6"$ west of Sgr~A* that is
free of obvious stellar sources.
The average positions and fluxes of Sgr~A* obtained from the NIR observations
are listed in Tab.\ref{SgrAdata}.
Fig.~\ref{Fig:Lightcurves} shows a plot of the flux
versus time for Sgr~A*, S1, and the background. As can be seen, the
background is fairly variable. These background fluctuations are
present in the light curves of Sgr~A* and S1 as well, as can be seen
in the Figure. It is therefore reasonable to subtract the background
flux from the light curves. The result is shown in Fig.~\ref{Fig:Flarefig1}. 
The source S1 now shows an almost constant flux density as expected.

\subsection{The Chandra X-ray Observations}
\label{section:Chandra}

In parallel to the NIR observations, SgrA* was observed with \emph{Chandra}
using the imaging array of the Advanced
CCD Imaging Spectrometer (ACIS-I; Weisskopf et al., 2002) for
two blocks of $\sim$50\,ks on 05--07 July 2004 (UT).  The start and
stop times are listed in Table~\ref{log}.
The instrument was operated in
timed exposure mode with detectors I0--3 turned on.  The time between
CCD frames was 
3.141~s.  
The event data were telemetered in faint
format.

We reduced and analyzed the data using CIAO v2.3\footnote{Chandra
Interactive Analysis of Observations (CIAO),
http://cxc.harvard.edu/ciao} software with Chandra CALDB
v2.22\footnote{http://cxc.harvard.edu/caldb}.  Following
Baganoff et al. (2003), we reprocessed the level~1 data to remove the
0.25\arcsec\ randomization of event positions applied during standard
pipeline processing and to retain events flagged as possible
cosmic-ray after-glows, since the strong diffuse emission in the
Galactic Center causes the algorithm to flag a significant fraction of
genuine X-rays.  The data were filtered on the standard ASCA grades.
The background was stable throughout the observation, and there were
no gaps in the telemetry.

The X-ray and optical positions of three Tycho-2 sources were
correlated (H{\o}g 2000) to register the ACIS field on the Hipparcos
coordinate frame to an accuracy of 0.10\arcsec\ (on axis); then we
measured the position of the X-ray source at Sgr~A*.  The X-ray position 
[$\alpha$$_{J2000.0}$ = $17^{\mathrm h}45^{\mathrm m}40.030^{\mathrm s}$,
$\delta$$_{J2000.0}$ = $-29\arcdeg00\arcmin28.23\arcsec$] 
is consistent with the radio position of Sgr~A* (Reid et al. 1999) to 
within $0.18\arcsec\ \pm 0.18\arcsec$ ($1\,\sigma$).

We extracted counts within radii of 0.5\arcsec, 1.0\arcsec, and
1.5\arcsec\ around Sgr~A* in the 2--8 keV band.  Background counts
were extracted from an annulus around Sgr~A* with inner and outer
radii of 2\arcsec\ and 10\arcsec, respectively, excluding regions
around discrete sources and bright structures
(Baganoff et al. 2003).  
The mean (total) count rates within the inner radius subdivided into 
the peak count rates during a flare and the corresponding IQ-values
are listed in Table~\ref{flareprop}.
The background rates have been
scaled to the area of the source region.  We note that the mean source
rate in the 1.5\arcsec\ aperture is consistent with the mean quiescent
source rates from previous observations 
(Baganoff et al. 2001, Baganoff et al. 2003).
The PSF encircled energy within each aperture increases from
$\approx50\%$ for the smallest radius to $\approx90\%$ for the
largest, while the estimated fraction of counts from the background
increases with radius from $\approx5\%$ to $\approx11\%$.  Thus, the
1.0\arcsec\ aperture provides the best compromise between maximizing
source signal and rejecting background.

\subsection{The VISIR MIR Observations}
\label{section:VISIR}

VISIR is the ESO mid-infrared combined imaging camera and spectrograph.
The camera is located at the  Cassegrain focus of UT3 (Melipal).
It operates over a wavelength range of  8-13 and 17-24 $\mu$m 
with fully reflective optics and two Si:As Blocked Impurity Band (BIB) 
array detectors from DRS Technologies. 
These detectors provide 256$\times$256 elements 
with a pixel scale of 0.075 arcsec/pixel for the 8-13$\mu$m wavelength range
and 0.127 arcsec/pixel for 17-24$\mu$m, resulting in fields of view of 
19.2"$\times$19.2" and 32.3"$\times$32.2", respectively.
The operating temperature is 15K and the detector temperature lies below 7K.
Background subtraction is performed via chopping and nodding.
The filter wheel is located just behind the cold stop pupil
and may hold up to 40 filters. 

The 8.6~$\mu$m and 19.5$\mu$m imaging data were taken on 
May 8, 9 and 10 (2004) as part of the
instrument commissioning. For the images used here, the 
larger scale of 0.127\arcsec/pixel was used. 
The angular resolution was of the order 0.3-0.4'' at 8.6$\mu$m and
0.6'' at 19.5$\mu$m
as measured from compact sources on the frames (e.g. IRS7).
The images each comprise
the average of typically seven chopped and flat-fielded exposures, with a
typical integration time of 14~s
(Fig. \ref{Fig:dustblobvisir}). 
Flux densities at the position of Sgr~A* and the S-cluster 
(combined, due to the resolution of 0.3-0.4'') were
extracted from these images, using IRS~21 as a flux calibrator. Tanner et al.
(2002) report 8.8~$\mu$m flux densities of 1.34~Jy (1\arcsec aperture,
background subtracted), 3.21~Jy (1\arcsec aperture, non-background-subtracted) and 3.6~Jy
(2\arcsec aperture, non-subtracted) for this source. 
After adjusting for the different
wavelength used (both wavelengths lie on the flank of the 9.7~$\mu$m silicate
absorption feature, c.f. Lutz et al. 1996), we obtain a background-subtracted
flux density of 5.11~Jy for IRS~21 in a 2\arcsec aperture. 
These flux densities also agree well with
the 8.7~$\mu$m flux densities for the bright IRS sources measured by Stolovy et
al. (1996), who found 5.6~Jy for IRS~21 (2\arcsec aperture, background not
subtracted). Recent results (Scoville et el. 2003, Viehmann et al. 2005) suggest a
lower extinction than the traditionally used $A_V$=27-30 magnitudes towards the
Galactic Center. Consequently, we adopt $A_V$=25 magnitudes and thus
$A_{\mathrm{8.6\mu m}}$=1.75 following the extinction law of 
Lutz et al. (1996),
which results in an extinction-corrected flux density of 4.56~Jy for
IRS~21 at 8.6~$\mu$m in a 1'' aperture, which we used 
for calibration.
The final flux density values at the position of SgrA* were obtained 
in a 1'' diameter aperture.
As reference sources to determine the relative positioning
between the NIR and MIR frames we used the bright and compact sources
IRS~3, 7, 21, 10W and in addition at 8.6$\mu$m wavelength IRS~9, 6E, and 29.
As a final positional uncertainty we obtain $\pm0.2''$.
At both wavelengths the flux density is dominated by an extended source 
centered about 0.2''-0.3'' west of SgrA*.
This source was also noted by Stolovy et al (1996).

In Tab.~\ref{VISIRfluxes} we list the flux densities obtained for the 
available 0.3-0.4'' angular resolution VISIR data.
We find a mean flux density of 30$\pm$10~mJy (1$\sigma$) at 8.6$\mu$m
wavelength with a tendency 
for larger flux density values during times of poorer seeing. 
The differences are also 
mainly between data sets taken at different days. There are no signs of a 
strong (i.e. several $\sigma$) deviation. We adopt a dereddened flux density 
at 8.6$\mu$m wavelength of 50~mJy as a safe upper limit.
Following the same calibration procedure, assuming  $A_V$=25, 
$A_{\mathrm{8.6\mu m}}$=1.75 and using 20.8$\mu$m flux densities for IRS~21
from Tanner et al. (2002), we find 
a flux density of 320$\pm$80~mJy at 19.5$\mu$m in a 1'' diameter aperture
at the position of SgrA*.

\subsection{The SMA Observations}
\label{section:SMA}
The Submillimeter Array observations
of SgrA* were made at 890~$\mu$m for three consecutive
nights, 05-07 July 2004 (UT), with the latter two nights timed to
coincide with the {\it Chandra} observing periods. These observations
were made with the Submillimeter Array\footnote{The Submillimeter
Array is a joint project between the Smithsonian Astrophysical
Observatory and the Academia Sinica Institute of Astronomy and
Astrophysics, and is funded by the Smithsonian Institution and the
Academia Sinica.} (SMA), an interferometer operating from 1.3~mm to
450~$\mu$m on Mauna Kea, Hawaii (Ho, Moran, \& Lo 2004). On both July 6 and
7 we obtained more than 6 hours of simultaneous X-ray/submillimeter
coverage (Table~\ref{log}); however, the longitudinal separation (nearly
90\arcdeg) between the VLT and the SMA limits the period of combined
IR/X-ray/submillimeter coverage to approximately 2.5 hours. The
weather degraded over the course of the three nights, from a zenith
opacity at 890~$\mu$m of 0.11 on July 5 to 0.15 and 0.29 on July 6 and
7, respectively. This is reflected in the larger time bins and scatter
in the later light curves.

The SgrA* observations were obtained as part of an ongoing SMA study
of the submillimeter polarization of this source (Marrone et al., in
preparation). Because these observations are obtained with circularly
polarized feeds, rather than the linearly polarized feeds usually
employed at these wavelengths, these data do not have the ambiguity
between total intensity variations and modulation from linear
polarization that afflict previous (sub)millimeter monitoring
(e.g. Zhao et al. 2003). This technique does mix the total intensity
with the source circular polarization, but measurements at 1.3~mm
(Bower et al. 2003, 2005) show no reliable circular polarization at
the percent level. In SMA polarimetric observations, only one out of
every 16 consecutive integrations on SgrA* is obtained with all
antennas observing the same polarization (left or right circular)
simultaneously, while the remaining integrations sample a combination
of aligned and crossed polarizations on each baseline. For the light
curves presented here 
(Figure~\ref{Fig:SMAlightcurve})
we use only the 50\% of integrations on each
baseline obtained with both antennas in the same polarization state,
as cross-polar integrations sample the linear polarization rather than
the total intensity. This unusual time sampling should not affect the
resulting light curves, as each point is an average of at least one 16
integration cycle. Nearby quasars were used for phase and gain
calibration. On July 5 and 6, the observing cycle was 3.5 minutes on
each of two quasars (1741$-$038, 1749$+$096), followed by 14 minutes
on SgrA*. On July 7 we used a shorter cycle and stronger quasars in
the poorer weather, 3.5 minutes on either 1741$-$038 or 1921$-$293
followed by 7 minutes on SgrA*, with the two quasars interleaved near
the end of the observations. Although 7 or 8 antennas were used in
each track, only the same 5 antennas with the best gain stability were
used to form light curves, resulting in a typical synthesized beam of
1\farcs5$\times$3\farcs0. The poor performance of the other antennas
can most likely be attributed to pointing errors.
We were concerned
about the effects of changing spatial sampling of smooth extended
emission on the flux densities we obtain from subsets of our SgrA*
observation. Images of the full three-day data set show very little
emission away from SgrA*, with peak emission around 3\% of the
amplitude of the central point source. To reduce sensitivity to the
largest angular scales we sample, we have excluded the two baselines
that project to less than 24~k$\lambda$ (angular scales $>$9\arcsec)
during the SgrA* observations. A comparison of the variations in the
three daily light curves versus hour angle shows no conclusive common
variations above the 10\% level, but the presence of differential
variations of 30\% or larger will mask smaller systematic changes. We
conclude that the surrounding emission is not responsible for changes
larger than around 10\%, and may in fact be significantly less
important. 

The SgrA* data is phase self-calibrated after the
application of the quasar gains to remove short-timescale phase
variations, then imaged and cleaned. Finally, the flux density is
extracted from a point source fit at the center of the image, with the
error taken from the noise in the residual image. The overall flux
scale is set by observations of Neptune, with an uncertainty of
approximately 25\%. The complete data sets from each night, including
other quasar observations not used in this analysis, show some
systematic day-to-day flux variations across sources at the 15\%
level. These have been removed here by assuming that the mean of the
quasar flux densities remain constant from day to day. Because we have
used multiple sources to remove this inter-day variation, we believe
this part of the absolute calibration to be accurate to about
10\%. Nevertheless, these uncertainties do not affect the errors
within each day's light curve.

\subsection{The VLA 7~mm Observations}
\label{section:VLA}

\begin{figure}
\centering 
\includegraphics[angle=0,width=13cm]{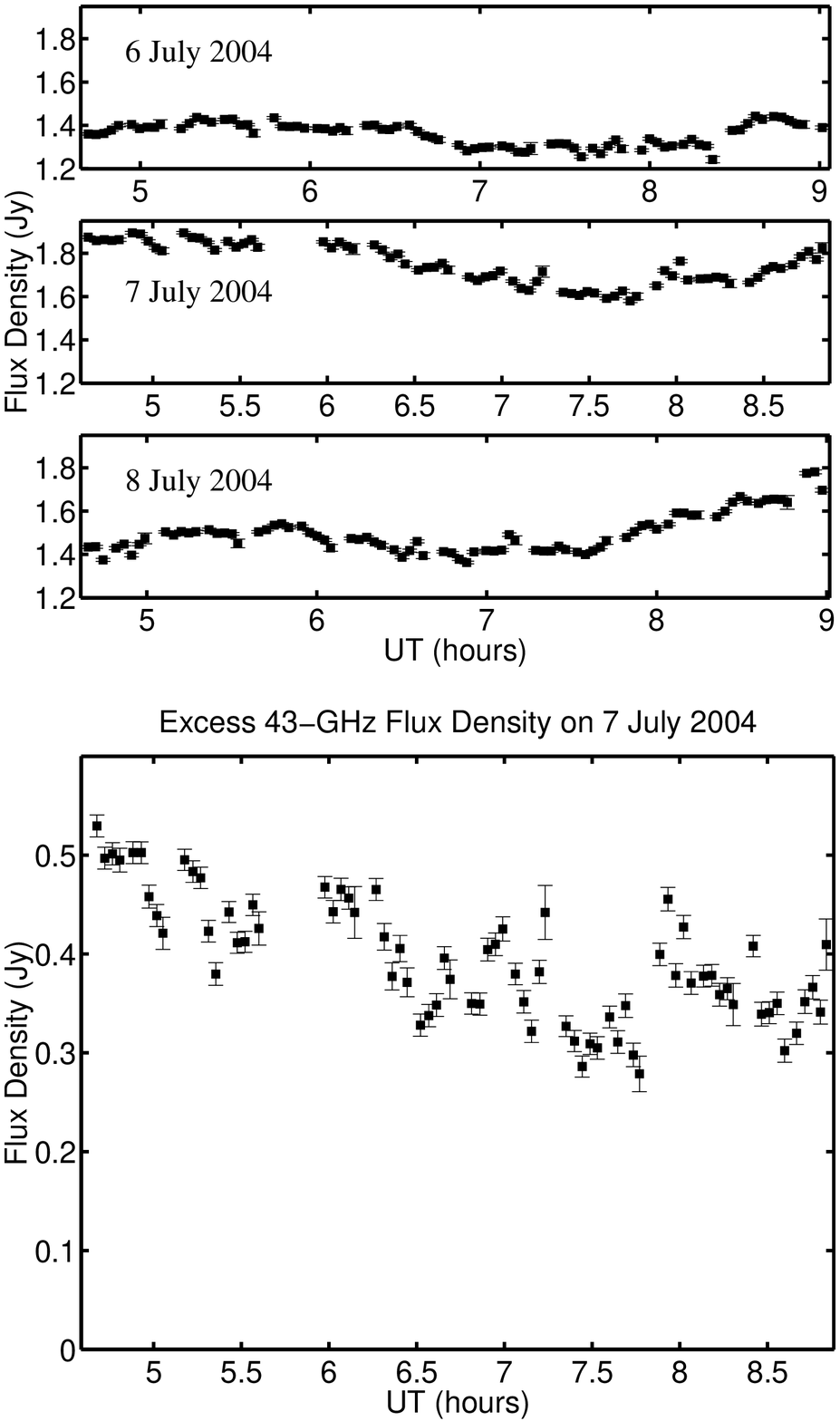}
\caption{\small VLA 43~GHz light curves. The top graphs show the
correlated flux density as measured on 6, 7, and 8 of July
(see Tab.~\ref{log}). The bottom graph shows the excess flux density
on 7 July calculated as the difference between the data from this 
day and the mean of 6 and 8 July.
}
\label{Fig:VLAlightcurve}
\end{figure}

The Very Large Array (VLA) 
observed Sgr A* for $\sim 5$ hours on 6, 7 and 8 July 2004. 
Observations covered roughly the UT time range 04:40 to 09:00 
(see also Table~\ref{log}), which is a subset of the Chandra observing
time on 6 and 7 July.  Observations on 6 July also overlapped with
NIR observations.
The VLA was in D configuration and achieved a resolution of $2.5 \times 0.9$
arcsec at the observing wavelength of 0.7 cm.  Fast-switching was
employed to eliminate short-term atmospheric phase fluctuations and provide
accurate short-term amplitude calibration.  For every 1.5 minutes on Sgr A*,
1 minute was spent on the nearby calibrator J1744-312.  Antenna-based
amplitude and phase gain solutions were obtained through self-calibration
of J1744-312 assuming a constant flux density.  Absolute
amplitude calibration was set by observations of 3C 286.  Flux densities
were determined for Sgr A* and J1744-312 through fitting of visibilities
at $(u,v)$ distances greater than 50 $k\lambda$.  This $(u,v)$-cutoff removed
contamination from extended structure in the Galactic Center.
We simultaneously fit for the flux density of the two components of a
transient source 2.7 arcsec South of Sgr A* (Bower et al. 2005).
The corresponding light curves of SgrA* for  6, 7 and 8 July 2004 are 
shown in Fig.~\ref{Fig:VLAlightcurve} (see also Table~\ref{log}).

\section{Properties of the Light Curves}
\label{section:NIRVariab}

\begin{figure}
\centering
\includegraphics[angle=270,width=15cm]{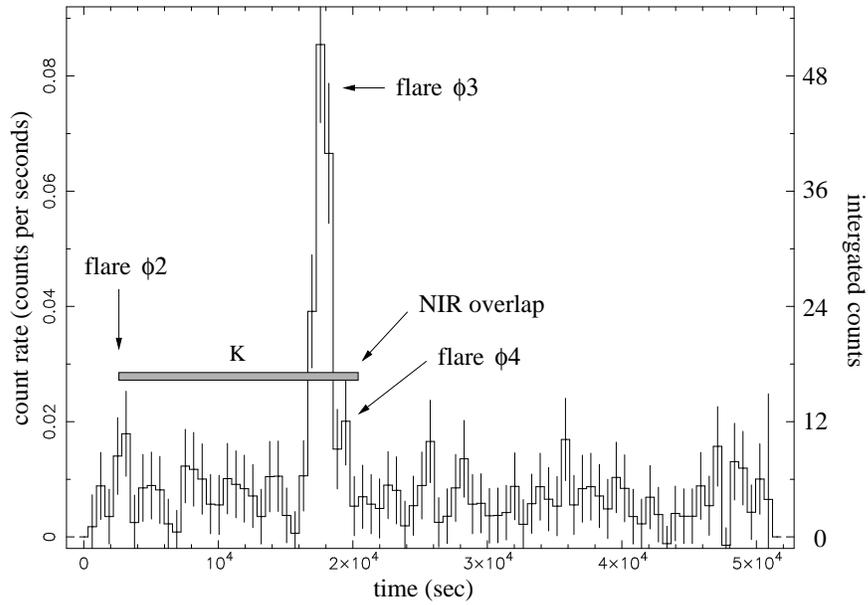}
\caption{\small \emph{Chandra} light curve for 6/7 July 2005.
We also indicate the overlap with the NIR data.
Start time is 2004 July 6, 22:35:11.8;
Stop  time is 2004 July 7, 12:53:44.9.
}
\label{Fig:chandra2}
\end{figure}

\begin{figure}
\centering
\includegraphics[angle=0,width=13cm]{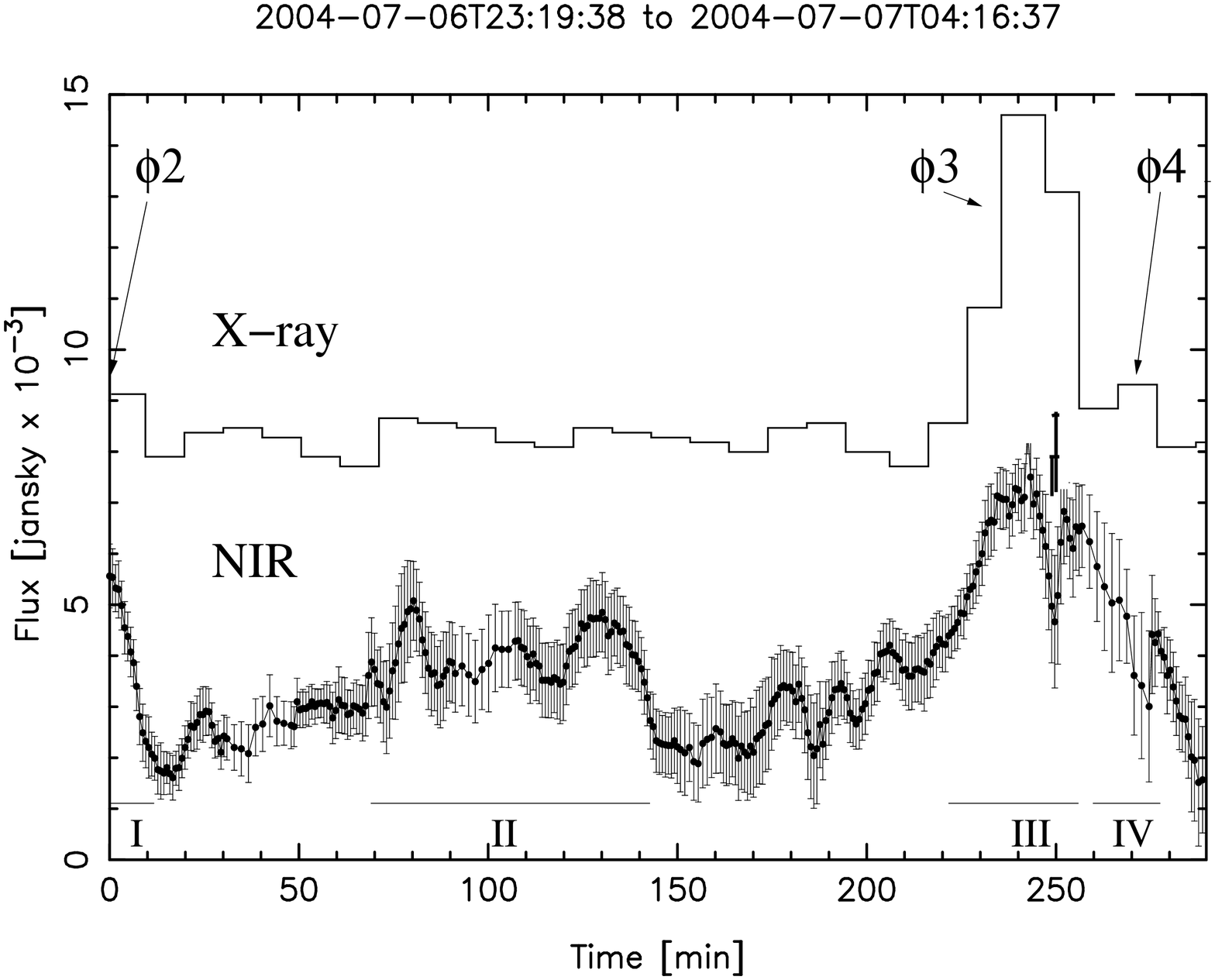}
\caption{\small The X-ray and NIR light curves plotted with a common
time axis.  See text and captions of previous figures.  Straight solid
lines in the inserted box represent the 0.00, 0.01, and 0.02 counts
per second levels.  The straight dashed line represent the X-ray
IQ-state flux density level.  
}
\label{Fig:nirxraylightcurves}
\end{figure}

The light curves show several flare events. In the following we 
base the identification of individual flares on their significant detection 
in at least one of the observed wavelength bands.
We have significant detections of a total of 5 NIR flare events 
(labeled I - V in Tab.~\ref{NIRobs} see also Fig.\ref{Fig:nirxraylightcurves})
and one bright X-ray flare 
(labeled $\phi$3 in Tab.~\ref{flareprop} see also Fig. \ref{Fig:chandra2}).
For 4 of the 5 NIR events we have 
simultaneous data coverage in both wavelength domains. 
In Figures~\ref{Fig:Flarefig1} and \ref{Fig:Flarefig2}, the 
light curve of the K-band imaging
from 7~July is shown separately. 
Four NIR bursts (I - IV) can be seen.
All of
them were covered by simultaneous \emph{Chandra} observations. In
Fig.~\ref{Fig:Flareims}, images corresponding to the time points
marked in Fig.~\ref{Fig:Flarefig2} are shown. The images show the
``on'' and ``off'' states of Sgr~A*.

\begin{figure}
\centering
\includegraphics[width=12cm]{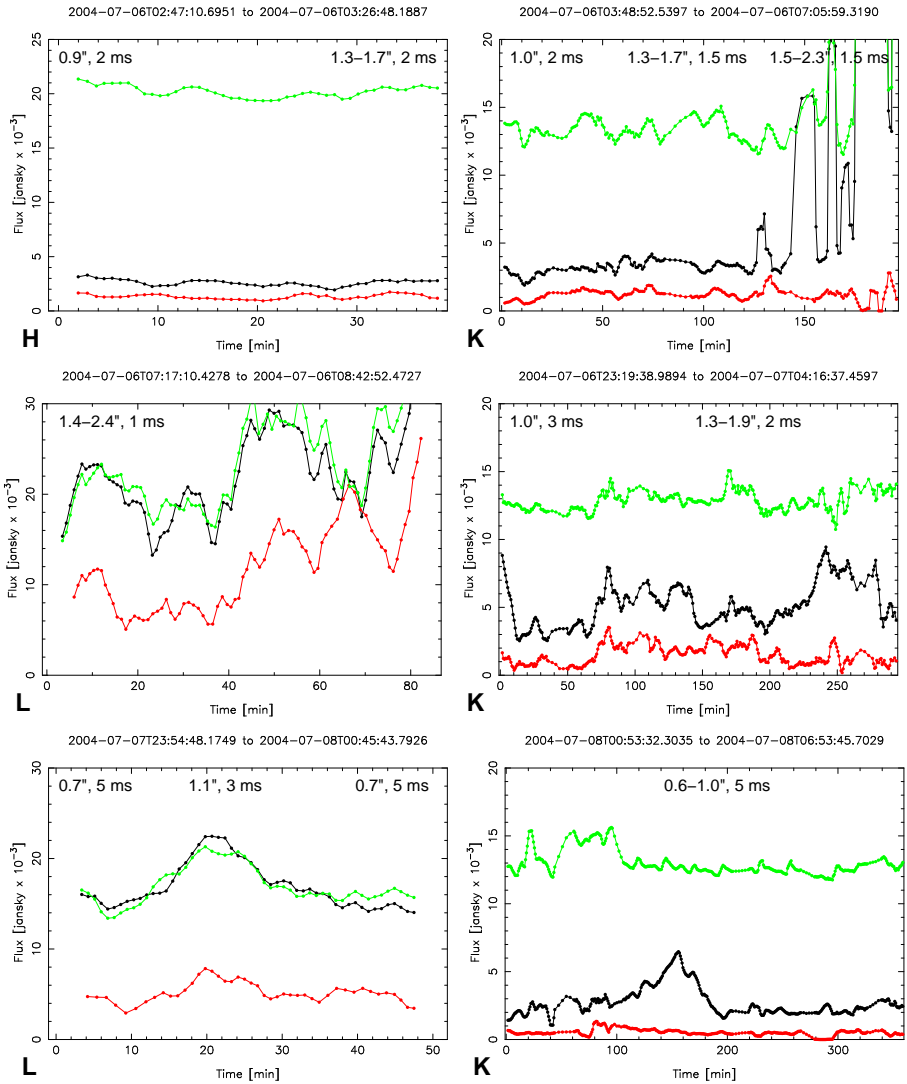}
\caption{Light curves of all NIR observations in July 2004. Shown are
  the dereddened flux densities of S1 (green), Sgr~A* (black), and a
  field free of stars (red). The data were smoothed with a sliding
  point window comprising five samples in case of the K- and H-band
  data, 9 samples in case of the L'-band data. Here, one sample
  corresponds to a photometric measurement on an individual image
  (i.e., DIT$\times$NDIT integration time, see Table~\ref{NIRObs}).
  Filters are indicated in the lower left corners, start and stop
  times on top of the plots. As can be seen, the quality of the data
  is fairly variable. This is closely related to the seeing and
  coherence time of the atmosphere during the observations, which
  result in a variable performance of the AO system. We indicated
  seeing and atmospheric coherence time in the plots. If there is more
  than one label in a given plot, then the location of the labels
  indicates approximately the time window to which the
  seeing/coherence time values apply. For example, in the L'-band data,
  seeing and coherence time deteriorated continuously during the
  observations on July 06/07, while in the July 08 data, only a brief
  episode of bad seeing occurred near the end of the first third of
  the time series. Generally, the light curve of S1 and of the
  background, which are expected to be constant, give a good idea of
  the data quality and relative uncertainties of the
  observations. Values for seeing and coherence time were taken from
  the web site of the ESO Observatories Ambient Conditions Database.
\label{Fig:Lightcurves}
}
\end{figure}

\begin{figure}
\centering
\includegraphics[width=12cm]{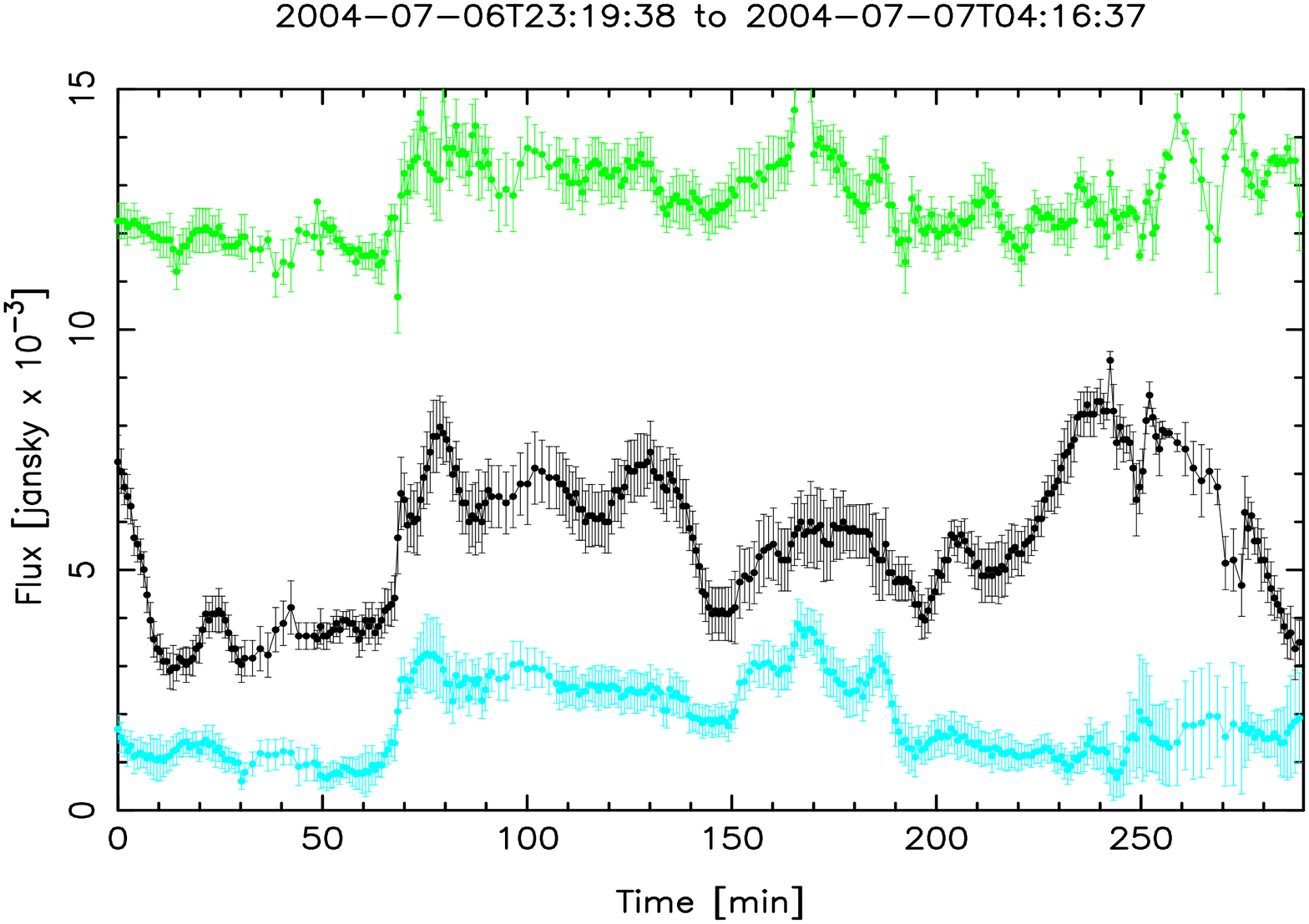}
\caption{Light curve of the K-band observations from July 07 (see
Fig.~\ref{Fig:Lightcurves}). The plot shows the lightcurves of 
S1 (top),SgrA* (middle), and
the background flux (bottom).
\label{Fig:Flarefig1}
}
\end{figure}

\begin{figure}
\centering
\includegraphics[angle=270,width=15cm]{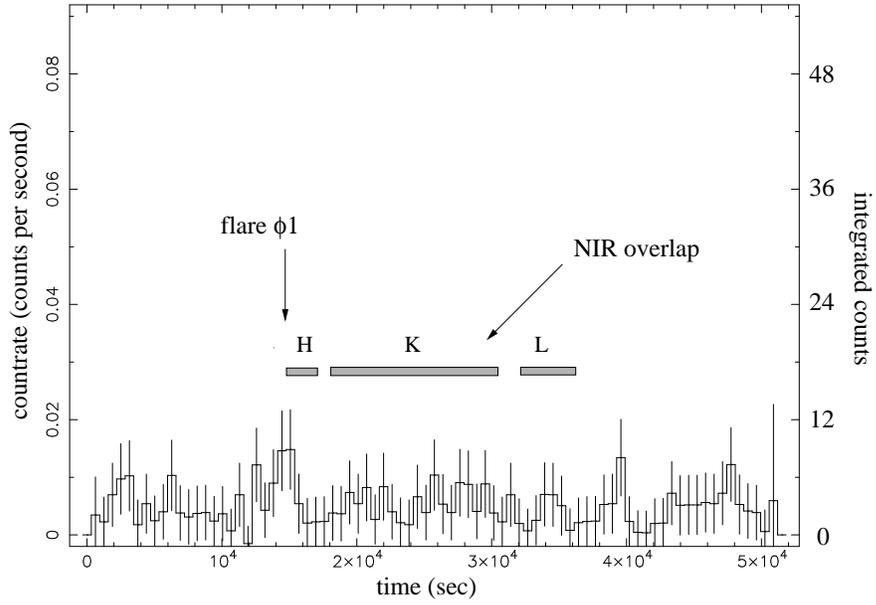}
\caption{\small \emph{Chandra} light curve for 5/6 July 2005.
We also indicate the overlap with the NIR data.
Start time is 2004 July 5, 22:38:25.7;
Stop  time is 2004 July 6, 12:56:58.8.
}
\label{Fig:chandra1}
\end{figure}

\begin{figure}
\centering
\includegraphics[width=12cm]{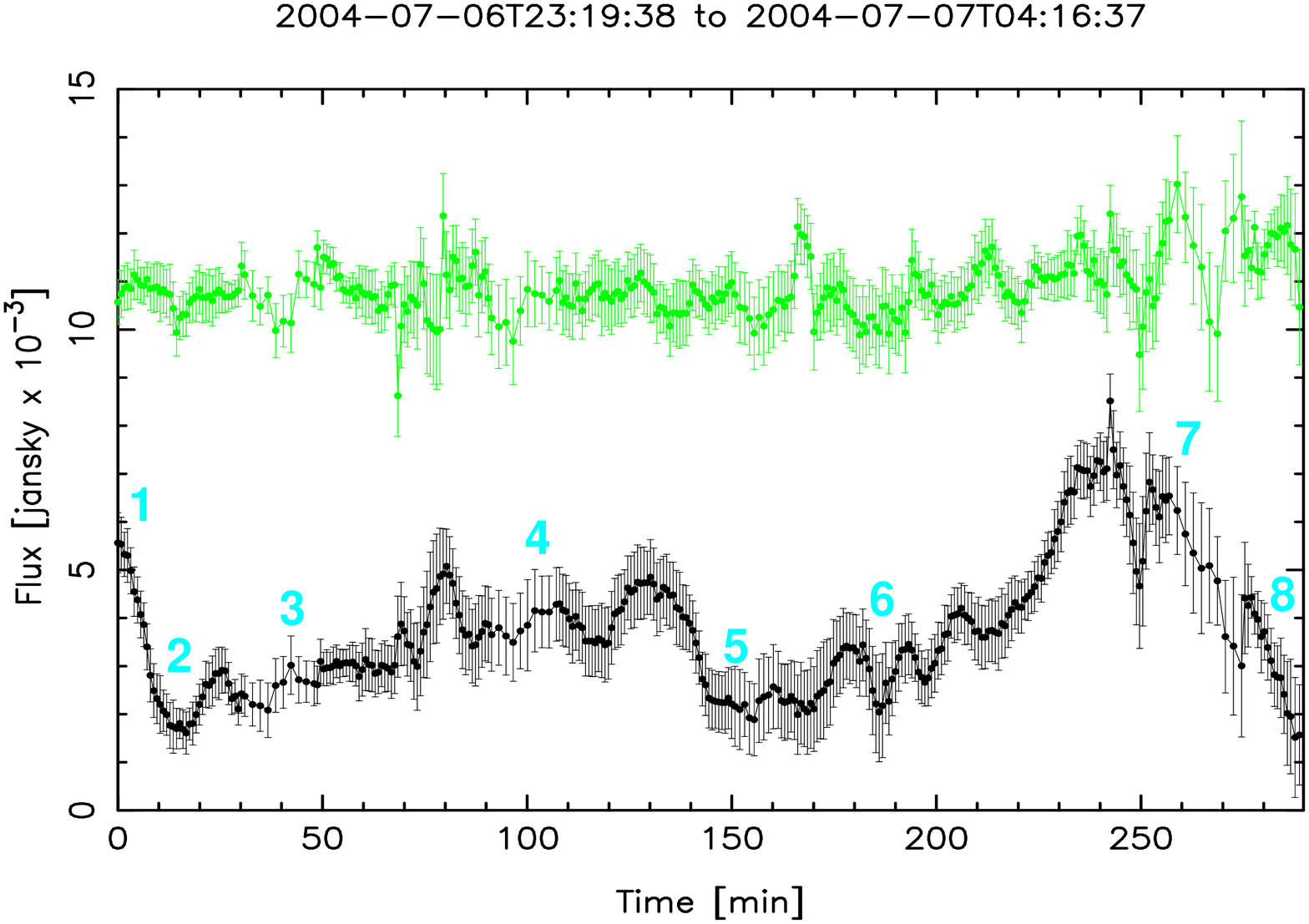}
\caption{Light curve of the K-band observations from July 07 (see
Fig.~\ref{Fig:Lightcurves}). 
The plot shows the lightcurves of S1 (top) and SgrA* (bottom). From both plots we 
have subtracted the background level.
The numbers near the light curve of Sgr~A*
mark the approximate time points for which images are shown in
Figure~\ref{Fig:Flareims}.
\label{Fig:Flarefig2}
}
\end{figure}

\begin{figure}
\centering
\includegraphics[width=12cm]{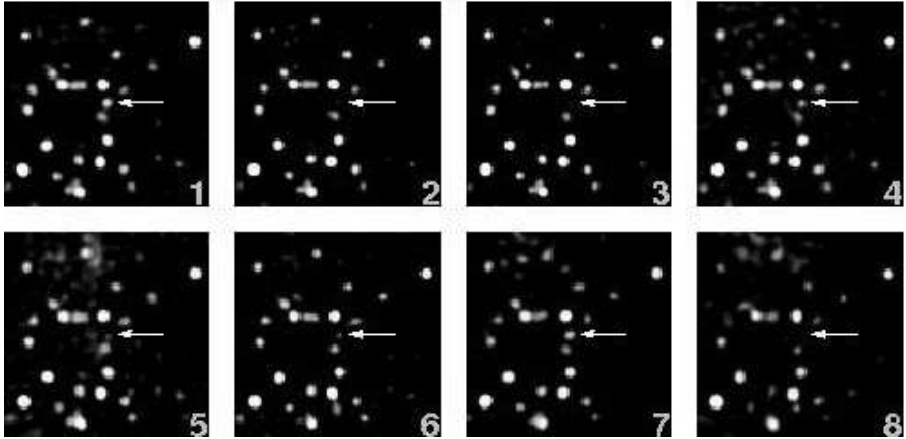}
\caption{K-band images of the stellar cluster in the immediate vicinity of
  Sgr~A*. The numbers correspond to the time points indicated in
  Figure~\ref{Fig:Flarefig2}. The images result from the average of
  five individual exposures, corresponding to 150\,s total integration
  time. A LR deconvolution and restoration 
  with a Gaussian beam was applied. 
  The color scale is linear. North is up, east to the
  left. The offsets are given with respect to the position of SgrA*.
  The white arrows indicate the position of Sgr~A*. 
\label{Fig:Flareims}
}
\end{figure}

\begin{figure}
\centering
\includegraphics[angle=0,width=13cm]{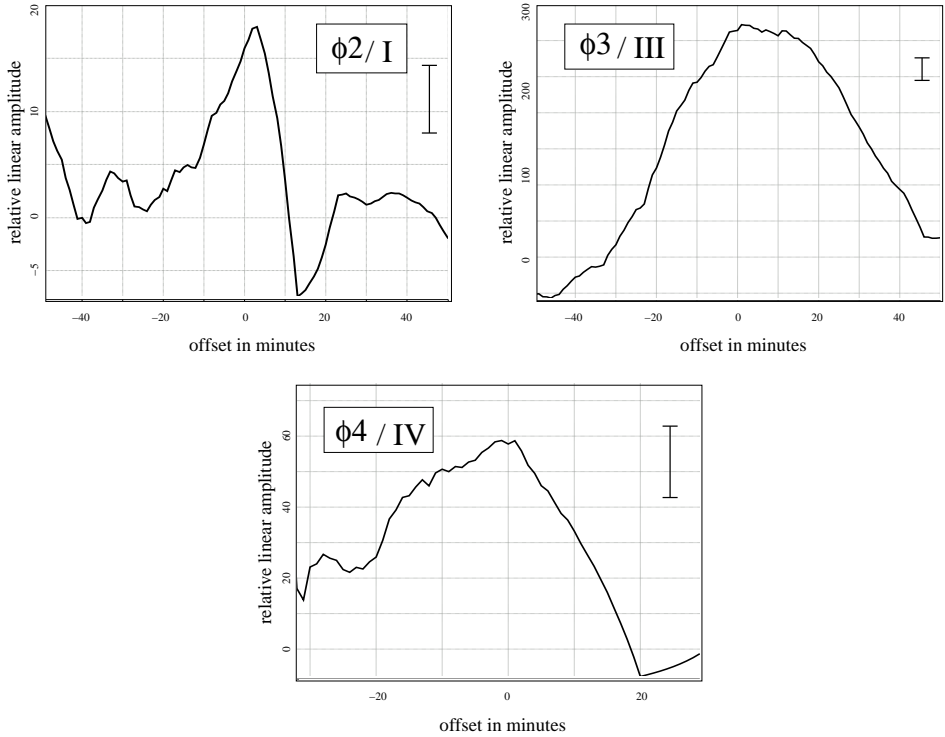}
\caption{\small Cross-correlation between the NIR data (40 second bins;
20 seconds integration time per image) and the X-ray data (10 minute bins)
for the three flares $\phi$2/I (top left),
$\phi3$/III (top right),
and 
$\phi4$/IV (bottom; see Tab.\ref{flares}).
We cross-correlated
only the flare data that overlap in time i.e. about 40 minutes before and 20 
minutes after midnight 07 July 0.0 UT for flare $\phi$2/I,
about 177 minutes until 223 minutes after midnight for $\phi3$/III,
and about 223 minutes until 256 minutes after midnight for $\phi4$/IV.
The error bars indicate an estimate of the SNR derived from the SNR of the 
individual X-ray bins and the approximate number of bins that were 
averaged for each of the cross-correlations measurements.
At a $\sim$2.6, 9, and 3$\sigma$ level, respectively, the light curves 
indicate a simultaneous flare
event around midnight corresponding to a time delay of less than 10
minutes (see text). 
For flares  $\phi3$/II  and $\phi4$/IV the entire event was covered.
The offsets are given in minutes. }
\label{Fig:correlation}
\end{figure}

In the X-ray regime we also draw attention to the weak flux density 
increases labeled 
$\phi$2 and $\phi$4 in Tab.~\ref{flareprop} 
(see also Fig. \ref{Fig:chandra1})
which coincide with significant NIR flares 
(labeled I and IV in Tab.~\ref{NIRobs} and Fig. \ref{Fig:chandra2}),
and to one weak X-ray flux density increase $\phi$1 which is similar
to $\phi$2.
In the following we consider these as weak flare candidates.
The interest in the weak events $\phi$2 and $\phi$4 is strengthened 
by the cross-correlation with the corresponding NIR flares (see Fig.~\ref{Fig:correlation}).
Fig.~\ref{Fig:chandra1} shows that the weak X-ray flare $\phi$1 did not have a
detectable  H-band counterpart. The H-band observations started almost 
exactly at the time of the peak X-ray emission.
As shown in Fig.~\ref{Fig:chandra2} the candidate (but insignificant by itself) 
weak X-ray flare $\phi$2 was covered
by NIR K-band measurements starting at its X-ray peak emission covering
the decaying part of the event. 
The excursion in the count rate labeled $\phi$2 
(which is insignificant by itself) occurs simultaneous to the significant 
NIR event II, which justifies its discussion.
The moderately bright X-ray flares $\phi$3 and $\phi$4 were fully covered 
in the NIR domain as well (see Figs.~\ref{Fig:chandra2} and \ref{Fig:nirxraylightcurves}). 
Flare $\phi$4 follows immediately after $\phi$3 and is similar in 
total strength and spectral index to $\phi$2.
The cross-correlation (Fig.~\ref{Fig:correlation})
between the X-ray and NIR flare emission
results in an upper limit for a  time lag between both events
of about 10 minutes.
The graph shows a clear
maximum close to 0 minutes delay indicating that within the binning
sizes both data sets are well correlated.  
Between the X-ray/NIR flares $\phi$2/I and $\phi$3/III we detect a 
lower level NIR flare phase which has no significant counterpart 
in the X-ray domain.
Finally the NIR flare V 
(see Figs.~\ref{Fig:Lightcurves} lower, right corner)
was not covered by our X-ray data.
In Tab.~\ref{flares} we give a summary of the observed flux densities and
spectral indices of all considered events.

We can estimate the times at
which the flare emission was negligible, i.e. equal to the 
low level variability IQ-state
emission in the X-ray and NIR domain.  
The corresponding full width at zero power (FWZP) and FWHM start
and stop times are given in Tabs.~\ref{flareprop} and \ref{NIRobs}.
For the weak candidate flare events $\phi$1, $\phi$2, and $\phi$4 
we only give estimates of FWZP.
In summary the statistical analysis of the combined X-ray and NIR data
for the 2004 observations shows that SgrA* underwent 
at least one significant flare event 
simultaneously in both wavelength regimes.

\begin{figure}
\centering 
\includegraphics[angle=0,width=14cm]{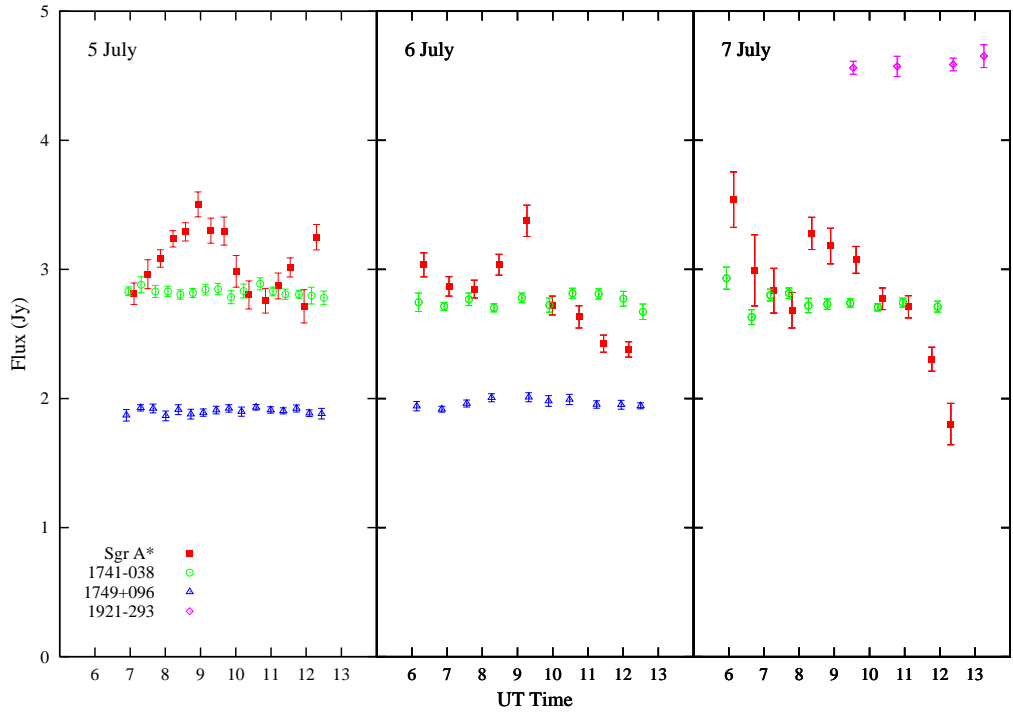}
\caption{The 890~$\mu$m light curves from July 05-07. The light curves of
the quasar calibrators, 1741$-$038, 1749$+$096, and 1921$-$293, are
also shown. For each source we plot Stokes I, the total intensity. Due
to poorer weather on July 06 and 07, for these days we have averaged
two cycles between SgrA* and the quasar calibrators (see text) for
each data point, while the July 05 data is shown at full time
resolution.}
\label{Fig:SMAlightcurve}
\end{figure}

The 890$\mu$m SMA submillimeter light curve 
(Figure~\ref{Fig:SMAlightcurve}) does not show an
obvious constant flux level, but instead is continuously varying
(as it is probably the case for the NIR emission as well). 
Several
flux density excursions of $10-20$\% are visible over the three
observing nights; the variations typically occur on $1-2$ hour
timescales, although there are also abrupt changes, such as the drop
around 09:30~UT on July 6 (SMA~3 in Tab. \ref{RADIOobs}).
These slow variations occur on somewhat longer
timescales than the X-ray and IR variations, perhaps suggesting that
the 890~$\mu$m emission extends out to regions tens of Schwarzschild
radii away, where causality slows the observed changes in the total
flux density integrated over the whole source. However, measurements
at 7~mm by Bower et al. (2004) suggest that the source is only
20$-$30~R$_s$ in diameter, corresponding to only about 10 minutes of light
travel time, so these slow variations are probably not due to
propagation effects.

The only period of coincident observations in the IR and
submillimeter, 2.5 hours on July 6, is rather featureless in the
submillimeter. Unfortunately, this portion of the IR observations is
not very reliable because very short atmospheric coherence times
resulted in poor AO performance (see Figure~\ref{Fig:Lightcurves},
upper right from minute 134 onward, and middle left). 

A comparison of the sub-millimeter and X-ray light curves
shows that on July 7 the decaying 43~GHz excess is accompanied by 
a $\sim$1~Jy sub-millimeter flux density decay
(SMA~4 in Tab. \ref{RADIOobs}).
After a small increase by about 0.5~Jy
(SMA~5 in Tab. \ref{RADIOobs})
that decay continues over the entire observing period on 7~July
(see red squares in the right panel of Fig.\ref{Fig:SMAlightcurve})
and amounts to a total decrease in flux density of about 2~Jy.
As the observing interval started about 2.3h after the
X-ray flare $\phi$3 and NIR flare III 
we speculate in section \ref{section:SSCmodel} 
that the sub-millimeter decay may be linked to 
adiabatic expansion of the emitting plasma.

Furthermore the comparison with the X-ray light curve
shows that one prominent submillimeter feature, a slow rise over 1.5
hours punctuated by a sharp drop in flux density at 09:30~UT on July 6,
(SMA~3 in Tab. \ref{RADIOobs})
is coincident with a very small increase in the X-ray flux at 39.4~ks in
Figure~\ref{Fig:chandra1}. 
This may be evidence that flares observed at shorter
wavelengths can be foretold by a slow rise in submillimeter
emission. This slow rise could manifest itself best in the
submillimeter if the IR and X-ray are dominated by SSC emission, which
varies non-linearly with the synchrotron emission. However, this
increase in the X-ray count rate is not very statistically
significant.

The 7~mm VLA radio light curves show relatively small variations on 6 and 8 July
with characteristic amplitudes of $\sim 10\%$.  On  July 7, however,
the flux density starts higher than on the previous day by about 0.5~Jy,
or $\sim 40\%$ of the mean flux density on  July 6 (Tab. \ref{RADIOobs}).
The flux density
declines steadily over the next three hours, reaching a minimum of 1.6~Jy.
The mean flux density on July 7 is $\sim$0.4~Jy in excess of the average
flux density on July 6 and 8.
The July 7 observations covered the time interval about
1.5 to 5 hours after the
the peak of X-ray flare $\phi$3 i.e. NIR flare III.  
Given the lack of simultaneous radio observations with the flare peak we can
only guess at the detailed relationship between the rise in the radio
flux density and the X-ray/NIR flare (see section \ref{section:SSCmodel}).
We cannot constrain whether
the radio flux density rise precedes or follows the X-ray flare, or,
for that matter, whether it is related to an X-ray flare that may have
occurred during the 10 hours between the two Chandra observations.

\section{NIR to X-Ray Flux Densities and Broad Band Spectral Indices}
\label{section:Fluxes}

Consistent with previous \emph{Chandra} observations
(Baganoff et al. 2001, 2003, Eckart et al. 2004)
the IQ-state X-ray count rate in a 1.5'' radius aperture  
during the monitoring period is
$5.3\pm0.5\times$10$^{-3}$~counts~s$^{-1}$.
This corresponds to a
2 - 8 keV luminosity of 2.2$\times$10$^{33}$~erg/s or a flux density
of $0.015\mathrm{\mu}$Jy. 
The excess X-ray flux density observed during the
strongest simultaneous flare
event $\phi$3 was 0.223$\pm$27~$\mu$Jy.  
This is about a factor of 15 higher than the IQ-state
and corresponds to a 2 - 8 keV
luminosity of about 33$\times$10$^{33}$~erg/s.  
For the other three events $\phi$1, $\phi$2, and $\phi$4 the excess was only 
a factor of 1.5 to 2 above the flux of the -state.
In the infrared the
$2.2\mathrm{\mu}$m extinction corrected flux density 
of the low level variability IQ-state 
Sgr~A* counterpart was 
found to be of the order of $3$~mJy and the excess flux density observed 
during the flares was only of the order of $5$~mJy.

For the stronger flare $\phi$3 
we find a spectral index of $\alpha_{X/NIR}$$\sim$1.12
(with $S_{\nu}$$\propto$$\nu^{-\alpha}$)
between the NIR regime (here at a
wavelength of $2.2 \mathrm{\mu}$m) and the X-ray domain (here centered 
approximately at an energy of 4~keV).
For the three weaker flares 
($\phi$1, $\phi$2, and $\phi$4)
we find $\alpha_{X/NIR}$$\sim$1.35, comparable to the value 
given for the event reported by Eckart et al. (2004)
(see also Tab.\ref{flares}).
This shows that the amplitude range of the X-ray flare emission is
larger than that observed in the infrared and that the infrared and X-ray
flare strengths are not necessarily proportional to each other.
It also suggests that stronger flare events may have a flatter overall spectrum.
The fact that the NIR spectral indices measured by 
Eisenhauer et al. (2005) and Ghez et al. (2005)
are usually much steeper than the overall X-ray/NIR spectral indices 
is discussed in detail in section
\ref{section:Interpretation}.


\section{The MIR/NIR Spectrum of SgrA*}
\label{section:Spectrum}

\subsection{The NIR Spectrum of SgrA*}

\begin{figure}
\centering
\includegraphics[width=12cm]{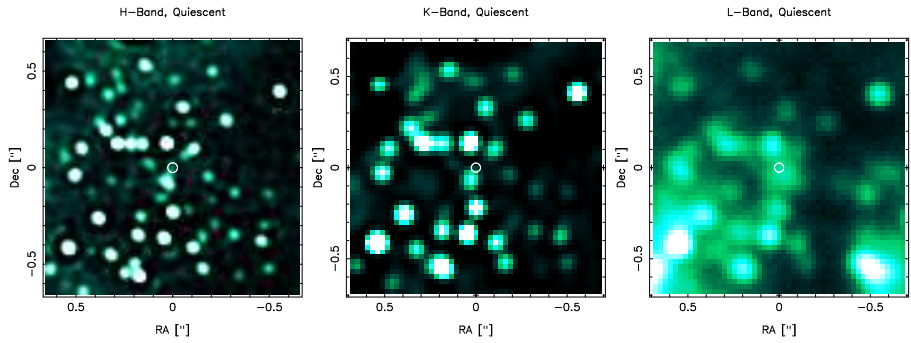}
\caption{Images of the stellar cluster in the immediate vicinity of
  the center during the IQ, low NIR flux density state of Sgr~A*. 
  The images result from the average of
  several individual exposures, corresponding to a total integration
  time of 40 minutes for H-band, 100 minutes for K-band, and 
  80 minutes for L'-band. 
  LR deconvolution and restoration with a Gaussian beam was
  applied. The color scale is linear. North is up, east to the
  left. The field of view shown in each image is $1.3''\times1.3''$.
  The white circles indicate the position of Sgr~A*. 
  All images were taken on the 6 July 2004. The positional
  accuracy of the images is of the order of 0.5 pixels, i.e.
  $\sim$7~mas in the H-band and $\sim$14~mas in the K- and 
  L'-bands.
\label{Fig:IQstate}
}
\end{figure}

On July 6 we covered the central region in the NIR H-, K-, and L'-band
with no detectable NIR flare emission during a quiescent
period with the exception of a short and weak decaying X-ray flare at
the beginning of our H-band exposure 
(NIR flare event I in Tab. \ref{NIRobs}).
In Fig.~\ref{Fig:IQstate} we show the corresponding
images during this phase in all three NIR bands (H-, K-, and L'-band). 
These images show that very little
emission originates from potential sources at the position of SgrA*. 
Both at H- and K-band the flux density may be severely influenced or even 
dominated by emission due to the stellar background.

The NIR spectral energy distribution of the flaring states of SgrA* appears to be very red. 
Using the new integral field spectrometer SINFONI at the 
VLT, Eisenhauer et al. (2005) obtained simultaneous
NIR spectral energy distributions ($\nu$L$_{\nu}$=$\nu^{-\epsilon}$)
of three flares.
They find that the slopes vary and have values ranging from 
$\epsilon$=2.2$\pm$0.3 to
$\epsilon$=3.7$\pm$0.9 
during weak flares of SgrA*.
This corresponds to a spectral index
$S_{\nu}$$\propto$$\nu^{-\alpha}$
range of about $\alpha$=3-5. 
Here we assume that their subtraction of spectral data 
cubes before the flare events
did only correct for scattered light/spillover and foreground/background 
stellar emission along the line of sight toward SgrA* and that the possible 
contribution of the flare emission to the flux density obtained at these times 
before the flare event is negligible.
Under these circumstances the differential spectral shapes can be 
regarded as true spectra of the NIR flares.

In recent narrow band measurements with the Keck telescope 
Ghez et al. (Ghez et al. 2005, and private communication) also 
found a red intrinsic flare spectrum.
These measurements, however, indicate a 
considerably flatter spectrum with a slope of $\alpha$=0.5$\pm$0.3.

\subsection{Contamination by MIR Dust Emission}

\begin{figure}
\centering 
\includegraphics[angle=0,width=14cm]{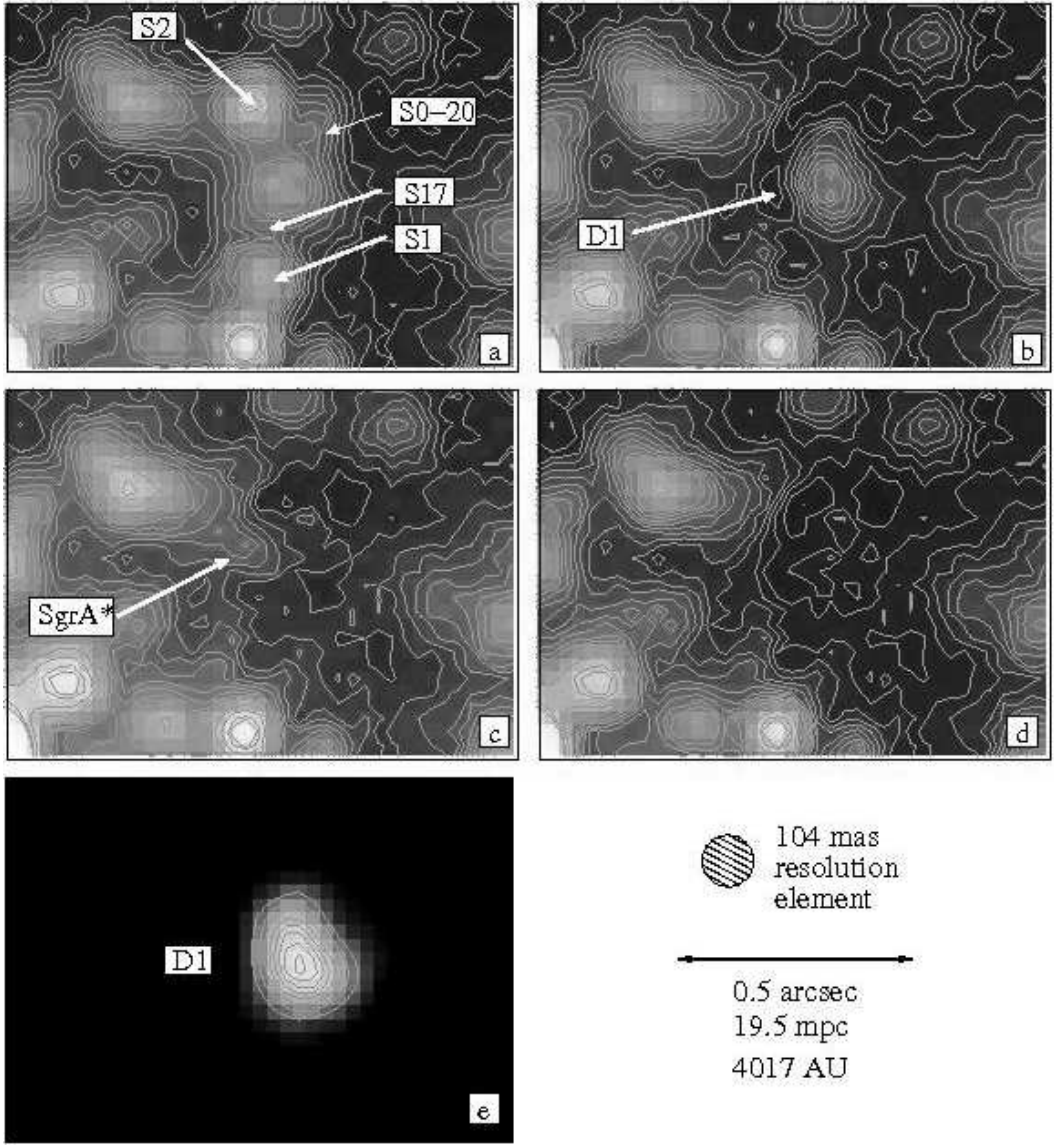}
\caption{\small 
The central 0.76''$\times$1.06'' shown in a color and contour map.
The contour lines are
4, 6, 8, 10, 12, 14, 16, 18, 20, 22, 24, 26, 30, 35, 40, 42
10, 14, 19, 24, 29, 33, 38, 43, 48, 52, 57, 62, 71, 83, 95, 
98 \% of the peak brightness of star S2.
Panel {\bf a)} shows the complete L' band images section.
Panel {\bf b)} shows the images section with scaled PSF subtractions
of the sources 
S2 (L'=12.78 used as a calibrator), 
S1 (L'=13.80$\pm$0.12), 
S17 (L'=13.79$\pm$0.12), 
S0-20 (L'=13.89$\pm$0.13), and 
SgrA* (L'=14.1$\pm$0.2). 
The dust component D1 (L'=12.8$\pm$0.20) is clearly visible.
Panel {\bf c)} shows the images with D1 subtracted instead of SgrA*.
In panel {\bf d)} both SgrA* and D1 have been subtracted.
Panel {\bf e)} shows the model of the dust component D1 by itself.
The background in the central 0.3'' is flat and has a L' brightness of
L'=15 per beam with a 3$\sigma$ deviation of less than $\pm$0.42 magnitudes.
}
\label{Fig:dustblobNACO}
\end{figure}

The images at wavelengths longer than 2.2$\mu$m 
show the existence of a
flux density contribution from a weak and extended source D1 with a separation
ranging between $\sim$30~mas and $\sim$150~mas from SgrA* (i.e. 80$\pm$50$mas$)
(see Fig.~\ref{Fig:dustblob} and 
\ref{Fig:dustblobvisir}; 
see also Ghez et al. 2004b, 2005).
The FWHM diameter of the source D1
is of the order of 
1400 AU at the distance of 8~kpc.  
In addition to the overall contribution of the mini-spiral 
to the dust emission there are several other red features of this 
kind labeled D2 to D8 
in Fig.~\ref{Fig:dustblob} within the 
central 2.6''$\times$2.6'' region.
At 8.6$\mu$m source D1
appears to be part of a larger ridge structure 
(tracing lower temperature dust; see also Stolovy, Hayward, \& Herter 1996)
at a position angle similar 
to a feature along an extension of dust emission from the mini-spiral towards
the central cluster of high velocity stars
(Fig.~\ref{Fig:dustblob}).
This ridge continues further to the northwest and we 
interpret this emission component close to SgrA* as 
part of this extension.
If this interpretation
is correct then this component is 
probably due to continuum emission of warm dust.
Deriving the flux density of this extended source D1 is difficult, as it is 
located in a crowded field and confused with the L'-band counterpart of SgrA*.
Using an L'-band magnitude of $m_{L'}$=12.78 for S2 
(Ghez et al. 2005, Clenet et al. 2005) and following the approach 
of Ghez et al. (2005) of subtracting all neighboring point sources, 
we can confirm the location, extent, and overall shape of this 
extended component (Fig.~\ref{Fig:dustblobNACO}).
Since no activity of SgrA* was indicated by the H and K band exposures 
before and the X-ray exposure during the L'-band imaging on July 6
(see Tab.\ref{log}) it is likely that SgrA* was in a low state.
On July 6 we find for SgrA* $m_{L'}$=14.1$\pm$0.2 (about 3.7 mJy dereddened),
which is
about half a magnitude brighter than what is reported by Ghez et al. (2005).
This is consistent with SgrA* having been in a low flux density state
during the L'-band exposure.
We also find that within the uncertainties (in subtracting the neighboring
sources including SgrA*) our data is consistent with 
an integrated L'-band magnitude of $m_{L'}$$\sim$12.8 for the extended
dust component, about 0.9 magnitudes brighter than what is given by
Ghez et al. (2005), i.e. the extended source D1 is almost as bright as star S2.
This would be consistent with a structure that is extended on scales larger
(see L' band magnitudes given in caption of Fig.~\ref{Fig:dustblobNACO})
than 100~mas, to which the VLT AO telescope beam couples slightly 
better than the Keck beam.
A brightness of m$_L'$=12.8$\pm$0.2 corresponds to a dereddened 
(m$_{ext}$=1.8) flux density of about 12$\pm$3mJy.

If the extended component D1 is associated with
a gas and dust feature of the Galactic Center ISM then the
8.6$\mu$m flux density limit is very likely contaminated by emission from
this extended dust feature as well. 
Assuming that this feature is not associated with SgrA* and has physical properties
similar to the other dust emission components in the central parsec, we can 
determine its contribution to the 8.6$\mu$m emission.
This can be approximated using its flux density value obtained in the L'-band 
and a mean flux density ratio between 8.6$\mu$m and 3.8$\mu$m obtained 
for the $\sim$200-400~K warm dust (Cotera et al. 1999).
From our available L'- and N-band images we derive this flux density ratio
of about 3 on
individual warmer sources (like IRS~21) and about 12$\pm$4
on the overall region (derived from our data and consistent with 
the ISO spectrum shown by Lutz et al. 1996)
which is dominated by the flux density 
contribution of the extended mini-spiral.
Hence we find from the L'-band flux density estimate of this component that 
within the uncertainties it can easily account for 
most of the 8.6$\mu$m flux density at the position of SgrA*. 
Similarly on the mini-spiral we measure a
flux density ratio between 19.5$\mu$m and  8.6$\mu$m of about
7$\pm$2 (consistent with Lutz et al. 1996). 
With the flux densities obtained at the position of SgrA* at 
8.6$\mu$m and 19.5$\mu$m the emission is consistent with being due to dust
similar to that found in the mini-spiral.

\begin{figure}
\centering
\includegraphics[angle=0,width=13cm]{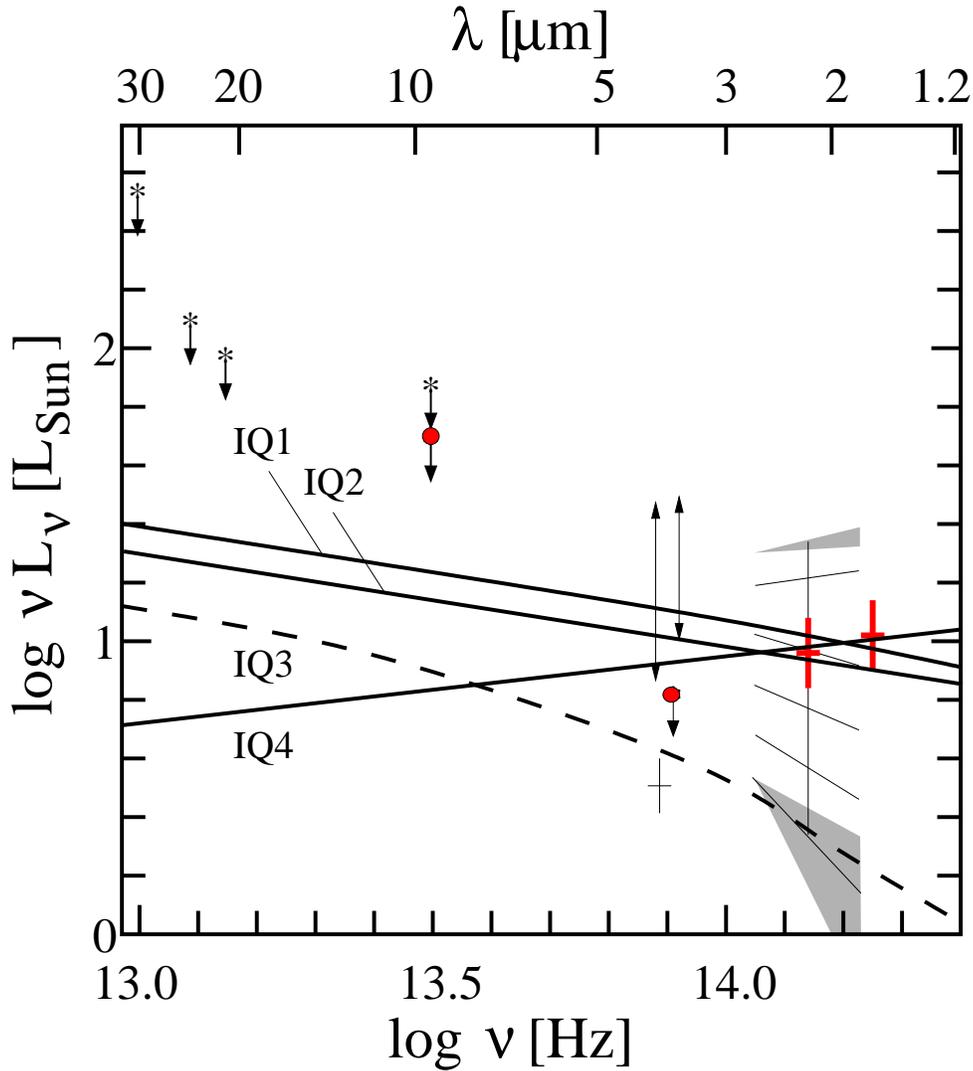}
\caption{\small The NIR/MIR spectrum of SgrA* compared to emission models
during low flux density states.
Red data points indicate the measurements described in this paper.
The log($\nu$ L$_{\nu}$) values are based on the following data:
Our new quasi-simultaneous L'-, K-, and H-band flux densities during 
the low NIR flux IQ-state are shown in red.
The 30$\mu$m and 24.3$\mu$m flux density limits are taken from
Telesco et al. (1996 APJ 456,541) and Serabyn et al. (1997),
respectively; at 8.7$\mu$m our limit of about 100~mJy is
lower than the value given by Stolovy et al. (1996); at 3.8$\mu$m 
our IQ flux density value (19$\pm$5 mJy) and the IQ flux density values
(6.4 - 17.5 mJy) given by 
Genzel et al. (2003) as well as the range of flux densities 
(4 - 17 mJy, 1.3$\pm$0.3 mJy) given by Ghez et al. (2004, 2005).
For clarity we displaced the latter two ranges slightly in 
frequency with respect to the center of the L'-band.
In addition we plot representative SSC models IQ1-IQ4 
with parameters as listed in Tab.~\ref{tab:models}.
}
\label{Fig:1NIRMIR}
\end{figure}

\begin{figure}
\centering
\includegraphics[angle=0,width=13cm]{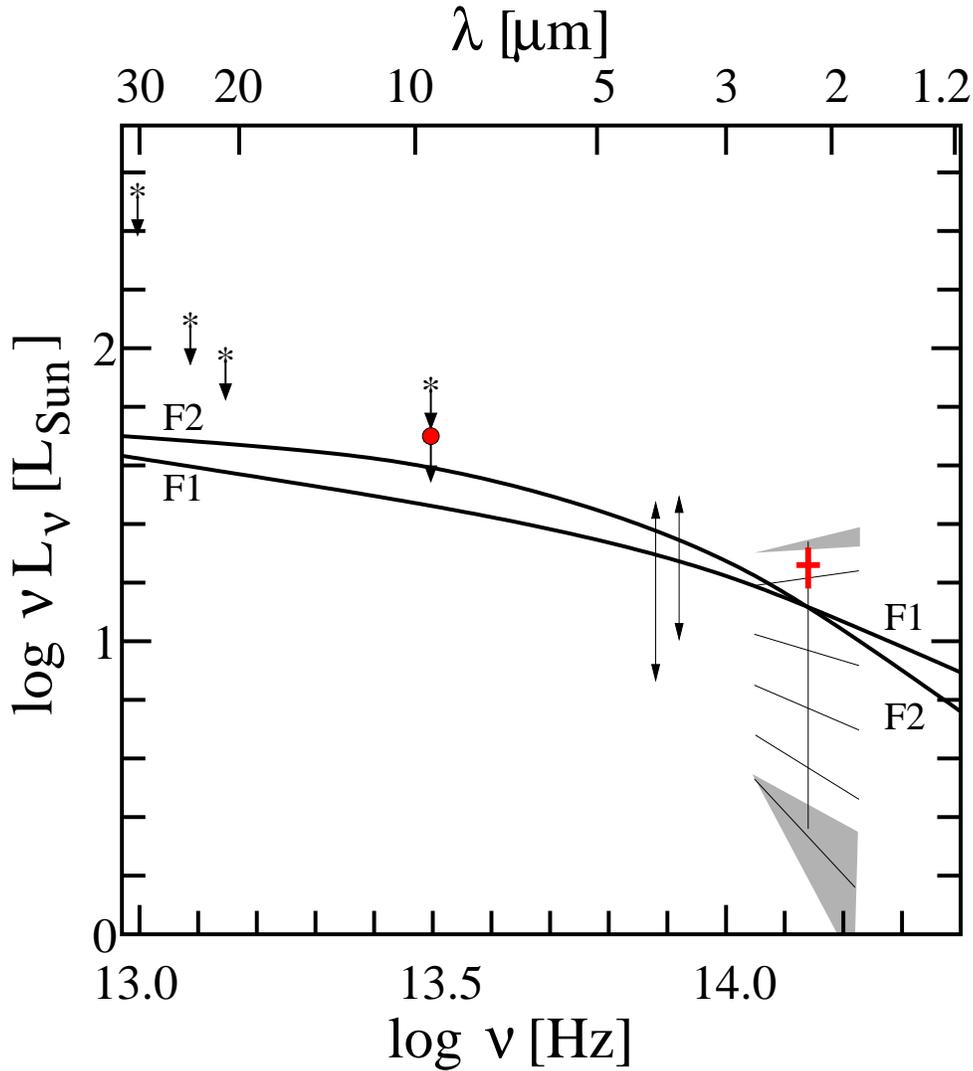}
\caption{\small The NIR/MIR log($\nu$ L$_{\nu}$) 
spectrum of SgrA* compared to flare emission models
for the strong X-ray/NIR flare $\phi~3/III$.
See also caption of Fig.~11.
Finally we show the K-band spectra of flares 
including the slopes (grey shaded area)
as measured by Eisenhauer et al. (2005) and Ghez et al. (2005).
The thin lines between the grey shaded areas 
indicate how the slope may change as a function of flare flux
as suggested by Ghez et al. (2005).
In addition we plot representative SSC flare models F1 and F2
with parameters as listed in 
Tab.~\ref{flares}.}
\label{Fig:2NIRMIR}
\end{figure}

In Figs.\ref{Fig:1NIRMIR} and \ref{Fig:2NIRMIR}
we show the flux densities or their limits towards SgrA*
expressed via the energy output $\nu$L$_{\nu}$ as a function of frequency $\nu$.
The plots cover the wavelength range between 30$\mu$m and 1.6$\mu$m.
In the mid-infrared ($\lambda$$\ge$8.6$\mu$m)  
only flux density limits are available.
For the NIR K-band we have plotted the envelope of the spectral
data  obtained by Eisenhauer et al. (2005) and Ghez et al. (2005).
For the L'-band we compare our value of the low level 
variability IQ flux densities with the corresponding
data intervals given by Genzel et al. (2003) and Ghez et al. (2004a, 2005) 
(see caption of Fig.\ref{Fig:1NIRMIR}).

It is likely that the L'-band IQ state continuum emission obtained 
for SgrA* is also to a small extent effected by 
a contribution from the weak, extended emission component 
(see section \ref{section:NIRVariab} and 
Fig.~\ref{Fig:dustblob} and \ref{Fig:dustblobvisir}.
This contribution will depend critically on the aperture used 
and the location of the 
red emission component with respect to SgrA*. We estimate that for the 
high angular resolution L'-band images this contribution cannot be larger than a few mJy.
For the 8.6$\mu$m band our flux density limit obtained from the
VISIR commissioning data is lower than the (dereddened) 
value of $\sim$100~mJy derived by 
Stolovy et al. (1996).
The values obtained by Telesco et al. (1996) and Serabyn et al. (1997) 
very likely include continuum emission contributions associated with
dust components in the central parsec.

Extrapolating the SINFONI spectral slopes towards lower frequencies,
their predicted 
flux densities lie within or above the quiescent state fluxes
obtained in the L'-band by 
Genzel et al. (2003), Ghez et al. (2004a) and the value reported in this paper.
Given that the data at longer wavelengths including those at 8.6$\mu$m 
only represent upper limits of the flux densities of SgrA*, the predicted 
values obtained by an extrapolation of the steep K-band spectra lie well 
above these limits and the expected intrinsic energy output of SgrA*.
This is especially true if a flux density level equivalent to the quiescent 
state were to be added back to the flare spectroscopy data.
Of course we do not know whether the MIR data have been taken during the
low level flux density (IQ)
 or flare state and in principle it is possible that SgrA* becomes very 
bright in the MIR during a flare. However, SgrA* has been observed frequently
during the past two decades and strong flare activity has never been 
reported at these wavelengths.
In total the combination of the very steep flare spectra and the low 
flux density limits of SgrA* especially at 8.6$\mu$m and longward imply 
that the intrinsic spectral energy distribution of 
SgrA* flattens significantly for wavelengths longward of 4$\mu$m.


\section{Flares in the NIR and X-Ray Domain}
\label{section:Flares}

The durations of flares found in the observations presented here
is in agreement with the current statistics.
Baganoff et al. (2001), Eckart et al. (2004), and Porquet et
al. (2003) report on X-ray events of 45 to 170 minutes.  
Eckart et al. (2004), Ghez et al. (2004a), and 
Genzel et al. (2003) report on NIR flare events that 
last 50 to 80 minutes, respectively.  
Simultaneous observations indicate that the NIR and X-ray flare events
are
well correlated in duration (see section \ref{section:NIRVariab}).
In the following we will assume as a working hypothesis that 
the activity of SgrA*
consists of consecutive flare events of variable strength
that have a characteristic duration of the order of 100 minutes. 
We further assume that flare event rate (the number of flares of a given
strength per day) and flare event strength can be described by a 
power-law.
Deviations from this assumption are discussed
towards the end of this section.

From the statistics of such flares and the NIR flux density monitoring 
that was compiled over the 
past decade one can assume a power-law representation that allows us
to predict the flare rate i.e. the number of flares of a given 
strength per day.
We assume that the power-law will be truncated at both ends. 
At the high end this can be justified by the fact that flares much stronger
than the neighboring high velocity S-stars have never been observed, and that 
the accretion process within a characteristic time scale must have a limited
radiation efficiency.
A truncation at the low end can be justified by the (currently) 
continuous supply of stellar wind material from the He-stars within the central
0.2~pc diameter of the stellar cluster.

Based on the measurements of 2003 the estimated infrared flaring rate is very
high: 4 IR flare events were found within a total of 25 hours of observations,
which results in about 2 to 6 events per day when assuming Poisson
statistics (Genzel et al. 2003).
With our most recent observations presented here we cover a total of
0.71 days and find 4 infrared flare events, 2 of which are above 5 mJy.
We can therefore confirm 
the rate of NIR (mostly) K-band 
flares of 4$\pm$2 per day with a strength of approximately 
10$\pm$5 mJy.
This leaves us with the rate of flares weaker than about 5~mJy of
10.4$\pm$2 per day (the remainder of 24~h not covered by $>$5~mJy 
flares divided by the characteristic flare duration). 
Here we include the IQ state and assume that it can be 
represented by weak, consecutive flares of the same average length
of 100 minutes.

\begin{figure}
\centering
\includegraphics[width=10cm]{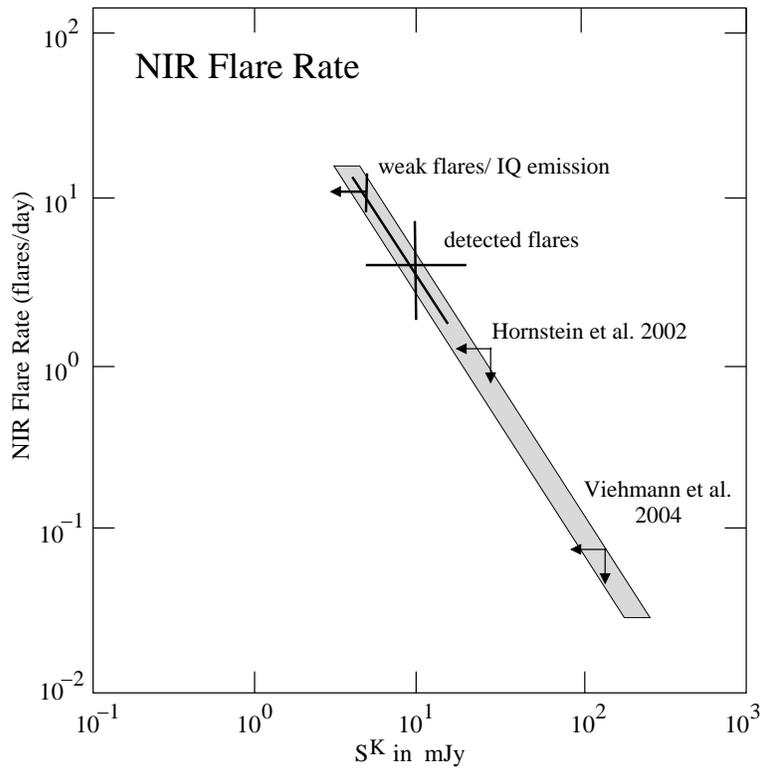}
\caption{\small 
 Flare amplitude as a function of flare rate for the NIR emission from
SgrA* under the assumption that the characteristic flare duration is
of the order of 100 minutes.  }
\label{Fig:flareplot}
\end{figure}

As for stronger flares, 
Hornstein et al. (2002) find from the analysis of Keck high angular
resolution  K-band imaging that the probability that a flare event of
3 hour duration has occurred with a flux density in excess of
19mJy is at most 9\%.
This corresponds to an equivalent flare rate of $\le$0.7 events of 3 hour duration
per 24 hours or $\le$1.3 per day assuming a flare duration of 100 minutes.
For the shorter duration we increase the detection  flux level by 
$\sqrt{2}$ to 27m~Jy.
Performing a similar analysis for ISAAC data Viehmann et al. (2004)
and Eckart et al. (2003) find a likelihood of 0.5\% for 3 hour flares with 
a flux density of more than 100mJy. The equivalent flare rates are 
3.6$\times$10$^{-2}$ for 3 hour flares and 
7.2$\times$10$^{-2}$ for 100 minute flares per day. 
Similarly we increase the detection  flux level by
$\sqrt{2}$ to 141~mJy
for the shorter assumed flare length of 100 minutes.
In Fig.~\ref{Fig:flareplot} we plot these quantities. 
The data can be described by a power-law of the form
$N=\kappa_0 (A)^{-\zeta} \kappa_1  \kappa_2$.
Here the quantities
$\kappa_0$ and $N$ are in units of flare events per 24 hours,
and  
$\kappa_1=exp(-\frac{A}{A_{high}})$
 and 
$\kappa_2=exp(-\frac{A_{low}}{A})$
allow for a truncation of the 
power-law towards strong and weak flare events.
$A_{high}$ and $A_{low}$ are cutoff flux densities at the 
high and low end of the power-law.
Treating the limits as real measurements we find $\zeta$=-1.4$\pm$0.2
and $\kappa_0$=100$\pm$30.
We note that in the case that our assumption that the flare length is independent
of the flare amplitude has to be modified, one can make the following 
statements:
If the overall characteristic flare duration is shorter or longer than 
100 minutes then the power-law will be largely unaffected. 
It would still pass through the point given by the detected flares. 
Since the flare rate would be higher and the flare fluxes 
lower, changing the flare duration would  correspond to a shift along the power-law line.
If the flare duration is shorter only for weaker flares then the power-law
can be extended towards the top.
If stronger flares become longer their rate goes down resulting in a
steeper slope of the power-law. Similarly if the flare duration is shorter
for stronger flares then the slope will be shallower.
An alternative to the truncated power-law representation that can currently 
not be ruled out is that the involved quantities like flare event rate, 
length, and strength are represented by peaked functions like a Gaussian.

It is highly improbable that the observed variability is due to
stellar sources because of the extremely short time scales of the
flares and because of the astrometric positions of the flares, which
are within less than 10~mas of Sgr~A* at all times 
(see also Tab. \ref{SgrAdata}):
a star close to
Sgr~A* would have moved by $\sim20-50$~mas during the time interval
covered by the four flares reported by 
Genzel et al. (2003).
A star at greater distances from Sgr~A* would have an extremely low
probability of being located so close in projection to Sgr~A*.

As for the relation between the NIR flares and variability in the
X-ray domain, the durations, rise, and decay times are similar
(see e.g. Baganoff et al. 2001, Porquet et al. 2003).
The NIR flare rate, however, was almost twice as high as the X-ray flare
rate during the \emph{Chandra} monitoring in 2002.
Also the range in spectral luminosities of the X-ray flares appears to be larger
than in the NIR (including the brightest X-ray flares).
X-ray flares a factor of $>$10 stronger than the quiescent emission occur at a rate
of $0.53\pm0.27$ per day, weaker flares are seen at a rate of $1.2\pm0.4$
(Baganoff et al. 2003).
Flares in the X-ray domain have been observed since 2000 
(Baganoff et al. 2001, 2003, Eckart et al. 2003, 2004, Porquet et al. 2003)
and since 2003 in the NIR domain 
(Genzel et al. 2003, Ghez at al. 2004a, Eckart et al. 2004).
Consequently further simultaneous observations are needed to determine the
relation between the X-ray and NIR flares.

\section{Physical Interpretation}
\label{section:Interpretation}

The new simultaneous X-ray/NIR flare detections of the SgrA* counterpart
presented here support the finding by Eckart et al. (2004) that
for the observed flare it is the same population
of electrons that is responsible for both the IR and the X-ray
emission.
Due to the short flare duration the flare emission very likely originates 
from compact source components.
The spectral energy distribution of SgrA* is currently explained by models
that invoke radiatively inefficient accretion flow processes (RIAFs: Quataert
2003, Yuan et al. 2002, Yuan, Quataert, \& Narayan 2003, 2004,
including advection dominated accretion flows (ADAF): Narayan et
al. 1995, convection dominated accretion flows (CDAF): Ball et
al. 2001, Quataert \& Gruzinov 2000, Narayan et al. 2002, Igumenshchev
2002, advection-dominated inflow-outflow solution (ADIOS): Blandford
\& Begelman 1999), jet models (Markoff et al. 2001), and Bondi-Hoyle
models (Melia \& Falcke 2001). 
Also combinations of models such as an accretion flow 
plus an outflow in form of a jet are considered 
(e.g. Yuan, Markoff, Falcke 2002).

\subsection{Description and Properties of the SSC Model}
\label{section:SSCmodel}

Current models 
(Markoff et al. 2001, Yuan, Markoff, Falcke (2002), Yuan, Quataert, \& Narayan 2003, 2004,
Liu, Melia \& Petrosian 2006)
predict that during a flare a few percent of the electrons
near the event horizon of the central black hole are accelerated.
These models give a description of the entire electromagnetic spectrum of SgrA*
from the radio to the X-ray domain.
In contrast we limit our analysis to modeling the NIR to X-ray spectrum of the most compact
source component at the location of SgrA*.
We have employed a simple SSC model to describe the observed 
radio to X-ray properties
of SgrA* using the nomenclature given by
Gould (1979) and Marscher (1983).
Inverse Compton scattering models provide an explanation for
both the compact NIR and X-ray emission
by up-scattering sub-mm-wavelength photons into these spectral domains.
Such models are considered
as a possibility in most of the recent modeling approaches
and may provide important insights into some fundamental
model requirements.
The models do not explain the entire low frequency radio spectrum and 
IQ state X-ray emission.
They give, however, a description of the compact IQ and flare emission 
originating from the immediate vicinity of the central black hole.
A more detailed explanation is also given by Eckart et al. (2004).

We assume a synchrotron source of angular extent $\theta$. 
The source size is of the order of a few Schwarzschild
radii R$_s$=2GM/c$^2$ with R$_s$$\sim$10$^{10}$~$m$ for a
3.6$\times$10$^6$\solm ~black hole. One R$_s$ then corresponds
to an angular diameter of $\sim$8~$\mu$as at a distance to the Galactic
Center of 8~kpc (Reid 1993, Eisenhauer et al. 2003).
The emitting source becomes optically thick at a frequency
$\nu_m$ with a flux density $S_m$, and has an optically thin spectral
index $\alpha$ following the law $S_{\nu}$$\propto$$\nu^{-\alpha}$.
This allows us to calculate the magnetic field strength $B$ and the
inverse Compton scattered flux density $S_{SSC}$ as a function of the
X-ray photon energy $E_{keV}$.  The synchrotron self-Compton spectrum
has the same spectral index as the synchrotron spectrum that is 
up-scattered 
i.e. $S_{SSC}$$\propto$$E_{keV}$$^{-\alpha}$, and is valid within the
limits $E_{min}$ and $E_{max}$ corresponding to the wavelengths
$\lambda_{max}$ and $\lambda_{min}$ (see Marscher et al. 1983 for
further details).
We find that Lorentz factors $\gamma_e$
for the emitting electrons of the order of 
typically 10$^3$ are required to produce a sufficient SSC flux in the
observed X-ray domain.
A possible relativistic bulk motion of the emitting source results 
in a Doppler
boosting factor $\delta$=$\Gamma$$^{-1}$(1-$\beta$cos$\phi$)$^{-1}$.
Here $\phi$ is the angle of the velocity vector to the line of sight,
$\beta$ the velocity $v$ in units of the speed of light $c$, and
Lorentz factor $\Gamma$=(1-$\beta$$^2$)$^{-1/2}$ for the bulk motion.
Relativistic bulk motion 
is not a necessity to produce sufficient SSC flux density but 
we have used modest values for 
$\Gamma$=1.2-2 and $\delta$ ranging between 1.3 and 2.0 (i.e. angles $\phi$ 
between about $10^{\circ}$ and $45^{\circ}$)
since they will occur
in case of relativistically orbiting gas as well as relativistic 
outflows - both of which are likely to be relevant in the case of 
SgrA*.

An additional feature of the model is that it allows to provide an estimate
of the extent of the pure synchrotron part of the spectrum by
giving the upper cutoff frequency $\nu_2$ of that spectrum as a function of 
source parameters including the maximum $\gamma_{e}$ of the relativistic electrons.
In order to explain the X-ray flare emission
by pure synchrotron models a high
energy cutoff in the electron energy distribution with large Lorentz
factors for the emitting electrons of $\gamma_e$$>$10$^5$ and magnetic
field strengths of the order of 10-100~G is required
(Baganoff et al. 2001, Markoff et al. 2001, Yuan, Quataert, \& Narayan 2004).
The correspondingly short cooling time scales of less than a
few hundred seconds would then require repeated injections or acceleration
of such energetic particles
(Baganoff et al. 2001, Markoff et al. 2001, Yuan, Quataert, \& Narayan 2004).
However, it is a more likely possibility that with $\gamma_e$$\sim$10$^3$, 
the cutoff frequency $\nu_2$ comes to lie within or just shortward 
of the NIR bands such that a considerable part of the NIR spectrum 
can be explained by synchrotron emission,
and the X-ray emission by inverse Compton emission.
This is supported by SSC models presented by Markoff
et al. (2001) and Yuan, Quataert, \& Narayan (2003) which result in a
significant amount of direct synchrotron emission in the infrared (see
also synchrotron models in Yuan, Quataert, \& Narayan 2004 and 
discussion in Eckart et al. 2004).

\subsection{Modeling Results}
\label{section:Modelingresults}

In the following we assume that the dominant  
sub-millimeter emitting source component 
responsible for the observed flares has a size
that is of the order of one to a few R$_s$ and a turnover 
frequency $\nu_m$ ranging from about 100~GHz to 1000~GHz.  
Eckart et al. (2004) have shown that at least the weaker flares 
can be described via a contribution of pure SSC emission 
both at NIR and X-ray wavelengths.
The corresponding magnetic field strengths are of the order of 0.3 to 40 Gauss,
which is within the range of magnetic fields expected for RIAF models
(e.g.  Markoff et al. 2001, Yuan, Quataert, Narayan 2003, 2004).
Also the required flux densities $S_m$ at the turnover frequency 
$\nu_m$ are well within the range of the observed 
variability of SgrA* in the mm-domain (Zhao et al. 2003, 2004).

Here we present in addition a few models which consist of a mixed 
contribution of synchrotron and SSC emission. With these models it is 
possible to describe both the low NIR/X-ray flux density state 
as well as the very red NIR flare spectra.

{\it The newly observed NIR/X-ray flares:}
While we do not have NIR in-band spectra of the flare events reported here, 
we have to take the available information on the K-band spectral indices into 
account.
Red and variable near infrared spectra are expected from most
model calculations (e.g. Markoff et al. 2001, Yuan, Quataert, \& Narayan 2004),
they are also compatible with the interpretation 
in the framework of a simple SSC model as described above
(see Eckart et al. 2004).
The fact that SgrA* apparently can have very red near infrared in-band
spectra during the flare phases
(Eisenhauer et al. 2005) combined with the low flux density limits
at wavelengths longward of the L'-band 
implies a significant spectral flattening of these very red intrinsic
SgrA* infrared flare events in the 4 to 10$\mu$m range
(This is of course under the assumption that the MIR flux densities are 
valid independent of whether or not SgrA* is in a flare or non-flare state.)
This suggests that a synchrotron component that experiences an exponential 
cutoff in the NIR/MIR wavelength range is responsible for a significant 
fraction of the flare state luminosity of SgrA*.

Two plausible, representative flare state models are listed in 
Tab.~\ref{tab:models} and are plotted in Fig.~\ref{Fig:2NIRMIR}.
For both models synchrotron radiation is produced up to frequencies of
200~THz i.e. just shortward of the NIR H-band.
We assume that this cutoff is due to an exponential cutoff
in the energy spectrum of the relativistic electrons. This
will represent itself as a modulation of 
the intrinsically flat spectra ($\alpha$=0.8-1.3)
with an exponential cutoff proportional to $exp[-(\nu/\nu_0)^{0.5}]$ 
(see e.g. Bregman 1985, and Bogdan \& Schlickeiser 1985)
and a cutoff frequency $\nu_0$ falling in the corresponding
wavelength range of 4-20$\mu$m.
Synchrotron losses at the high end of the
relativistic electron spectrum may be responsible for such a cutoff.
Small variations in such an exponential damping of the radiation provide
{\bf variable,} red flare spectra.
Within the uncertainties models like F1 or F2 reproduce the NIR/X-Ray 
properties of the observed 
flare $\phi$1/III (Tab.~\ref{tab:models}) very well.
Model F1 or the pure SSC model presented in Eckart et al. (2004) may
represent flare $\phi$2 and $\phi$4.
Flare $\phi$1, which has not been detected in the H-band, is consistent with
flare $\phi$2 and $\phi$4 and the indication that the flare emission is 
intrinsically very red (Eisenhauer et al. 2005).

Ghez et al. (2005) suggests that the NIR spectral index is a function of
NIR flare brightness with weak flares ($\sim$2mJy or less at 2.2$\mu$m -
comparable to $\phi$1) 
having steep intrinsic NIR spectra ($\alpha$$\sim$4)
and brighter flares ($\ge$6mJy at 2.2$\mu$m comparable to $\phi$3)
having flatter intrinsic NIR spectra ($\alpha$$\sim$0.5).
In this context an exponential cutoff would be required for 
weak flares, where as for brighter flares the intrinsic spectral indices 
in the models (see Tab.~\ref{tab:models}) are much closer to the
spectral index derived by Ghez et al. (2005).

Our model results show that intrinsically very flat 
($\alpha$$\le$0.5) synchrotron spectra 
result in a large discrepancy between the measured and 
predicted X-ray and NIR flux densities
and large magnetic fields
unless a spectral cutoff in the 50-100$\mu$m range is introduced 
which makes the spectra significantly steeper again in the NIR.
Smaller source sizes and higher turnover frequencies 
$\nu_m$ of a few 1000~GHz result in very large ($>$100 G) magnetic fields
as well.

{\it The low flux density IQ-state of SgrA*:}
Over the central 0.6 arcsecond radius 
the X-ray flux density is due to extended 
thermal bremsstrahlung from the outer regions of an accretion flow
(R$>$10$^3$ R$_s$; Baganoff et al. 2001, 2003, see also Quataert 2003).
We find that the compact X-ray emission in the 'interim-quiescent' (IQ), 
low-level flux density states of SgrA* can be explained by 
a SSC model that allows for substantial contributions from both the 
SSC and the synchrotron part of the modeled spectrum.
In these models the X-ray emission of the point source is well below 
20$-$30~nJy and contributes much less than half of the X-ray flux density 
during the weak flare event reported by Eckart et al. (2004).
The flux densities at a wavelength of 2.2$\mu$m are of the order
of the observed value of of 1 to 3 mJy during the IQ-state which 
is in full agreement with a state of low level flux density variations.
Representative models for the low flux state are listed in 
Tab.~\ref{tab:models} and plotted in Fig.~\ref{Fig:1NIRMIR}.
For models IQ1-IQ3 
the source component has a size of the order of 1 to 2 Schwarzschild radii
with an optically thin radio/sub-mm spectral index ranging from
$\alpha$$_{NIR/X-ray}$$\sim$1.0 to 
$\alpha$$_{NIR/X-ray}$$\sim$1.3, 
a value similar to the
observed value between the NIR and X-ray domain.

For models IQ1 and IQ2 (see Tab.  \ref{tab:models})
the upper cutoff frequency  $\nu_2$ of the synchrotron spectrum lies just 
within or short of the observed NIR bands.
Here the SSC IQ models represent lower bounds to the 
measured flux density limits at the position of SgrA*.
Model IQ3 shows that a longwavelength cut-off in the MIR
results in the steep NIR spectra that have been observed
by Eisenhauer et al. (2005) and Ghez et al. (2005).
Model IQ3 is also in agreement with the low L'-band 
flux densities reported by Ghez et al. (2005) and in this paper.

Model IQ4 shows that rather blue spactra 
may also be a possibility for the low flux state.
In model IQ4 in which no NIR/X-ray cutoff is involved,
the flare radiation originates predominantly from a
synchrotron component that is smaller than 
fraction of a Schwarzschild radius.

{\it Modeling the mm- and sub-mm radio data:}
The observations of simultaneous NIR and X-ray flare emission suggests
a flare source size of the order of one Schwarzschild radius R$_s$, 
whereas the measured source size at radio wavelengths 
is of the order of 20$-$30~R$_s$ or 160$-$240~$\mu$as at 43~GHz (Bower et al. 2004).  
While it cannnot be excluded that this is purely due to some opacity structure that
makes it much larger at longer wavelengths, it may also be 
consistent with the assumption that 
the source responsible for the NIR
emission expands as it cools.
Such a scenario is actually supported by the (quasi-)simultaneous 
millimeter to X-ray observations of the bright flare emission presented here.
If we assume that the X-ray flare $\phi$3 and NIR flare III are 
physically associated with the decaying radio and sub-millimeter
flux density excess 
detected 1-2 hours later with the VLA and SMA, 
the corresponding radio decay 
timescale of a few hours and its amplitude are
factors that a consistent model (like the one presented below) should account for.

The models in Table 9 produce very little ($<$ 10 mJy at 43 GHz) instantaneous 
flux density at frequencies below the peak frequency of the flare. 
Motivated by the overall spectral shape of SgrA*,
we assume a THz peaked flare model like F1 or F2, 
assume a self-absorbed synchrotron spectrum at lower frequencies,
and adiabatic
expansion of the synchrotron emitting flare component via~~
$S(\nu_2)=S(\nu_1)[\nu_1/\nu_2]^{-(7\gamma_e+3)/(4\gamma_e+6)}$~
(van der Laan 1966).
The radio flux density will then first rise and later drop 
as the source evolves as indicated by the radio data following 
$\phi3$ and $\phi4$ on July 7 
(in comparison to the radio data on the previous day).
Here  $\gamma_e$ is given via the spectral index of the optically thin part
of the synchrotron spectrum~~ $\alpha=(\gamma_e-1)/2$.
Then with $\alpha$$\sim$-1.0 the peak flux density of S($\nu_1$)=11~Jy 
at $\nu_1$=1.0-1.6~THz will relate to a peak flux density 
of about S($\nu_2$)= 0.1-0.2~Jy at $\nu_2$=43~GHz and 
S($\nu_2$)= 1.5-2.6~Jy at $\nu_2$=340~GHz (890$\mu$m).
At $\nu_2$=340~GHz the flux density should then drop by 0.5 Jy, which is 
very comparable to what has been observed with the SMA.
These flux density contributions represent a major portion of the 
observed radio and sub-millimeter excess emission on 7 July.
Adiabatic expansion would also result in a slower decay rate and 
a longer flare timescale at lower frequencies, as it is observed (less than an 
hour at NIR/X-ray wavelengths and more than 3.6 hours in the radio domain).

As the flare expands it will cool on the synchrotron cooling time scale. 
This can be calculated via~~
$t_s \sim 3 \times 10^7 \Gamma \delta^{0.5} \nu_9^{-0.5} B^{-3/2}$,
where $t_s$ is in seconds, B is in Gauss, $\nu_9$ is frequency in GHz, 
and $\gamma$ and $\Gamma$ are the relativistic factors for the bulk 
motion of the material (Blandford \& K\"onigl 1979).
The synchrotron cooling time at 1.6~THz, 340~GHz, and 43~GHz for $B=68~G$ with 
$\Gamma$$\sim$$\delta$$\sim$1.5 
is about 1, 1.8, and 4 hours, respectively.
This matches well to the observed flux densities and decay time scale.

If we assume that the emission originates from relativistically orbiting material, then
the source may expand starting as a compact 
8.5 $\mu$as radius source over the entire orbit with a $\sim$200~$\mu$as radius.  
This will most likely happen at a velocity close to sound speed $c/\sqrt(3)$
(Blandford \& McKee 1977), implying a time scale of about 2 hours.
If we think of the emitting source as a component in a freely expanding jet
the expansion from  a 8.5 $\mu$as to 240 $\mu$as radius source at
a speed $c/\sqrt(3)$  will happen in about 14~minutes.
Within the given set of assumptions the model of relativistically
orbiting material gives a more suitable representation of the
observed flux densities and decay time scale - \underline{unless} the
possible jet is foreshortened since it is pointed at the observer.

The comparison of model F1 and F2 with the NIR data also indicates that
the very steep spectral slopes found by
Eisenhauer et al. (2005) are most likely linked to flare events that do 
not produce a significant amount of SSC radiation in the NIR.
These flares (probably due to a less energetic population of relativistic
electrons) would then be dominated by the exponentially decaying 
direct synchrotron component rather than a contribution of inverse Compton 
radiation (due to a more energetic population of relativistic
electrons).


\section{Summary and Discussion}

We have presented new, successful simultaneous X-ray and NIR
observations of SgrA* in a flaring and the IQ low NIR flux  density state.  
We found 4 X-ray flares (2 definite flares and 2 putative events) and
5 NIR flares with  4 events covered simultaneously at both wavelengths.
For the flares we observed simultaneously in both wavelength domains, the 
time lag between the flares at different wavelengths 
is less than 10 minutes and therefore consistent with zero.
Combined with the information that the NIR flare spectra are very red
with variable spectral indices (Eisenhauer et al. 2005, 
Ghez et al. private communication) we can successfully 
describe the flares by a SSC model in which a substantial fraction 
of the NIR emission is due to a truncated synchrotron spectrum.
Inverse Compton scattering of the THz-peaked flare spectrum by the 
relativistic electrons accounts for the X-ray emission.

Our investigation also shows that the NIR K-band is the ideal wavelength band 
to study the flare emission from SgrA*. In combination with adaptive optics 
systems it provides the highest angular resolution at the lowest amount of 
contamination by dust emission. 
At wavelengths shorter than the K-band little emission is
found because the flares are red (Eisenhauer et al. 2005) and 
at longer wavelength 
the angular resolution is lower and the dust contamination is high.
Observations in the K-band allow us to measure the highest 
flare rate and are - in the framework of the presented physical model -
ideally suited to observe both synchrotron and SSC flare emission.
In addition the model also gives us the opportunity to perform polarization
measurements which could provide additional information to study the 
relevant emission mechanisms.

The total number $\Sigma$ of detectable flares can be obtained by integrating 
over the amplitude dependent flare rate 
$N(A)=\kappa_0 (A)^{-\zeta} \kappa_1  \kappa_2$ 
(see section \ref{section:Flares})
 as 
$\Sigma=\int_{A_{limit}}^{\infty}N(A)dA$, with $A_{limit}$ being the 
detection limit of the flare emission. 
The model presented in section~\ref{section:Interpretation}
suggests that in the NIR domain the observed flares can be produced by
a mixture of synchrotron and SSC emission, i.e. 
$\Sigma_{NIR}=\Sigma_{Synch,SSC}$.
Since we can assume that the X-ray flares are predominantly produced by
SSC emission rather than synchrotron emission, as also suggested by the
very steep NIR flare spectra (Eisenhauer et al. 2005), it follows that
$\Sigma_{X-ray}\sim\Sigma_{SSC}$. 
As a consequence - and in good agreement with the observations - 
the total number of detected X-ray flares is smaller than that in the NIR 
$\Sigma_{X-ray} \le \Sigma_{NIR}$. This does, however, not imply that 
the flux density distribution of flares dominated by SSC emission 
in both wavelength domains are the same.
That distribution depends on the 
properties of the relativistic electron spectrum responsible
for the emission at both wavelength regimes. 
These are reflected
in parameters like the spectral index of the optically thin radio continuum
and the exact location of the high and low energy cutoff frequencies of the
scattered SSC spectrum. In addition, NIR flares may have contributions 
from both the synchrotron and the SSC part the flare spectrum and it may be 
difficult to discriminate between the SSC and synchrotron dominated flare activity.
One can, however, expect that SSC dominated NIR flares are bluer 
than synchrotron dominated ones.

\begin{figure}
\centering
\includegraphics[width=10cm]{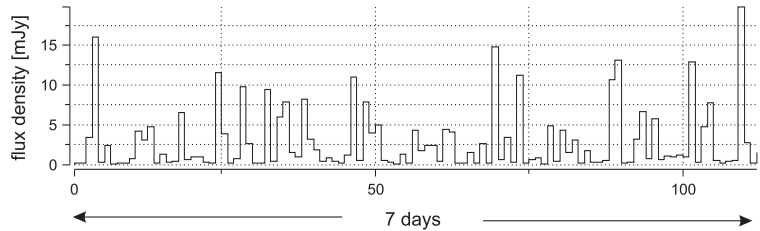}
\caption{\small 
 Simulation of the Sgr~A* flare
activity assuming a power spectrum relation between flare amplitude
and the number of flares. Here the power-law spectrum is comparable to 
the value found from our NIR flare data and each bin covers one 
characteristic flare time.}
\label{Fig:nirsim}
\end{figure}

The description of the flare activity as a power-law under the assumption of a 
characteristic flare time implies that the IQ phase can be regarded as 
a sequence of frequent low amplitude flares of SgrA*. 
Such a model would predict phases
of very low flux densities (see also Eckart et al. 2004 - Garching). 
In Fig. \ref{Fig:nirsim} we show a simulation with the appropriate power-law spectral index.
The observed flares may be the consequence of a clumpy or turbulent accretion.
Evidence of a hot turbulent accretion flow onto SgrA* based on polarization 
measurements has been discussed by Bower et al. (2005).
In this case the flare power spectrum is coupled to the power spectrum of
accreted clumps or the turbulences in the accretion flow.

\begin{figure}
\centering 
\includegraphics[angle=0,width=14cm]{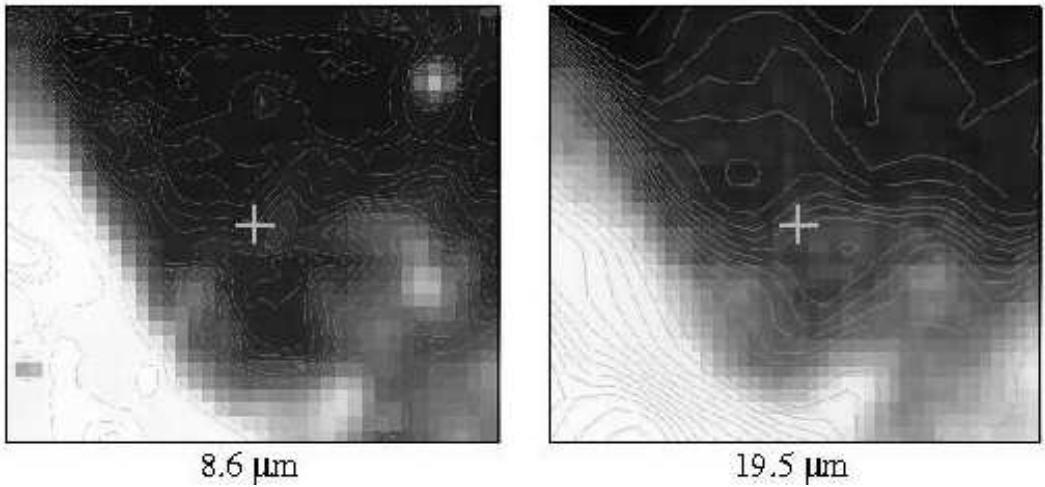}
\caption{\small 
The 8.6$\mu$m (left) and 19.5$\mu$m (right)
image of the central $5.0''\times4.4''$ as taken with VISIR
on 8 May 2004.
The emission at the position of SgrA* is consistent with dust
emission similar to what is measured towards other faint components
within the central 5-10 arcseconds.
The cross marks the position of SgrA* and the positional uncertainty
of $\pm0.2''$. As reference sources to determine the relative positioning
between the NIR and MIR frame we used the bright and compact sources
IRS~3, 7, 21, 10W and, in addition at 8.6$\mu$m wavelength, IRS~9, 6E, and 29.
}
\label{Fig:dustblobvisir}
\end{figure}

\begin{figure}
\centering 
\includegraphics[angle=0,width=14cm]{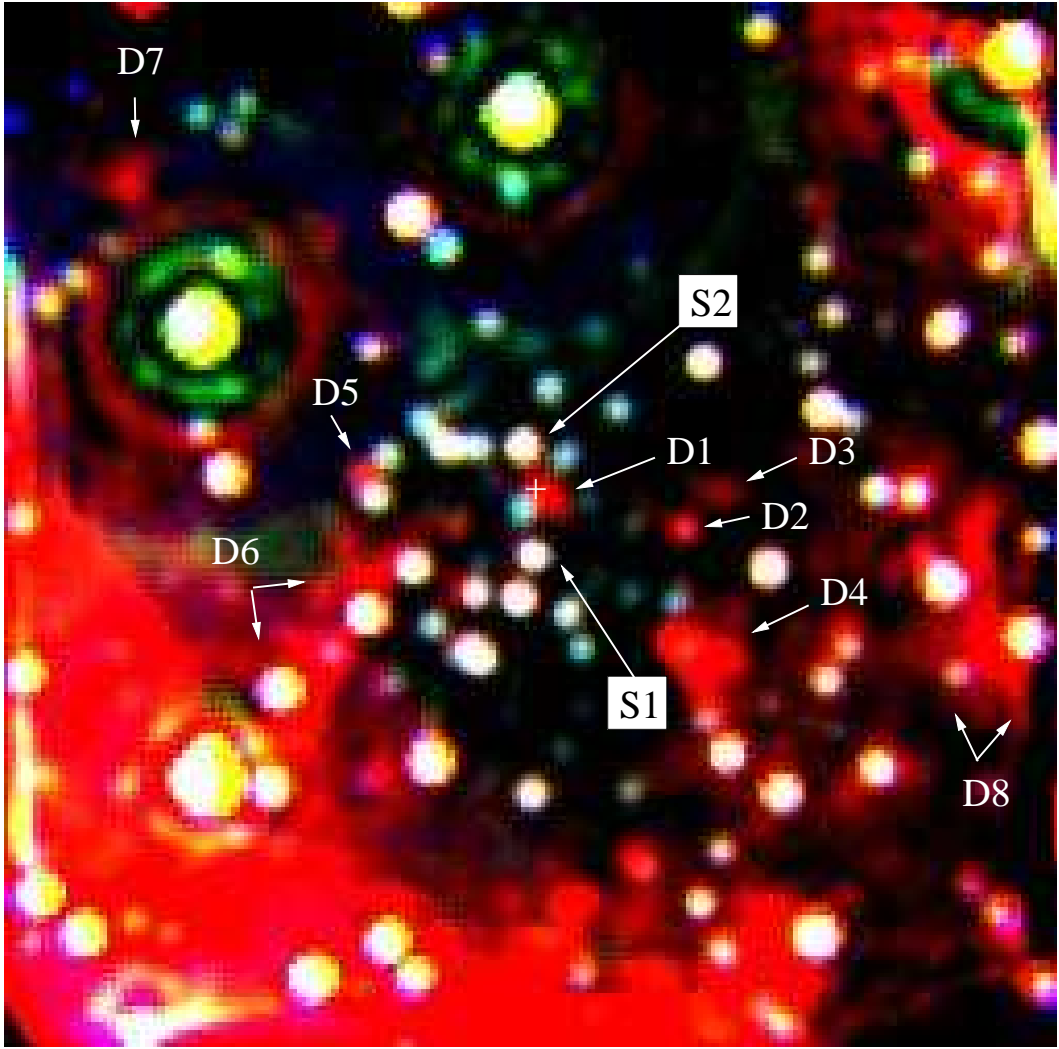}
\caption{\small HKL' multi-color image of the central $2.6''\times2.6''$
as taken with NACO on 6~July 2004. North is up and East is to the
left. H is blue, K is green, and L' is red.
The white cross marks the position of SgrA*.
The figure is a different representation of the same data shown in 
Fig.~\ref{Fig:IQstate} covering a larger field of view.
Just west of the position of SgrA* an extended red emission 
component D1 is evident.
}
\label{Fig:dustblob}
\end{figure}

The red source component we identified close to the position of SgrA*
at 3.8$\mu$m, 8.6$\mu$m, and 19.5$\mu$m is probably contaminated significantly
by thermal emission from a dust component along the line of sight towards
SgrA* (Fig.~\ref{Fig:dustblob} and \ref{Fig:dustblobvisir}).
The infrared flux density ratios of the emission from that region compared to
values obtained from the mini-spiral and other discrete sources in the
central parsec suggest that the emission is due to dust.
Assuming that the gas and dust properties of this component are similar to
the material in the northern arm we can obtain a first order estimate 
of its mass, which can be thought of as a structure which is thin 
with respect to its projected extent (e.g. Vollmer \& Duschl 2000).
Based on CO(7-6) measurements Stacey et al. (2004) derive a total gas mass of 
the northern arm of 5 to 50 \solm. In projected size the dust component 
close to 
SgrA* covers about 1/250 of the areas comprised by the northern arm.
This results in a gas mass of the order of 10$^{-2}$\solm
(If the dust temperature is substantially higher than $\sim$200-400~K
which are typical of the mini-spiral - see Cotera et al. 1999 -
then the overall mass of this component can be considerably 
smaller; see Ghez et al. 2005).

Depending of the clumpiness of the gas distribution within that component
on the source size scale of SgrA* this may result in a significant 
column density.
The dust source is, however, most likely located behind SgrA*.
If it were located in front of SgrA* the high velocity stars in the central 
cusp would also be affected. 
However, for other sources in the field, like S2, S12, and S14, their 
(A$_V$$\sim$25$^m$) extinction corrected spectra are blue, and variable 
flux densities and colors
have not been detected within the uncertainties of a few 0.1 magnitudes.
The variable NIR spectral indices reported for the red flares 
(Eisenhauer et al. 2005, Ghez et al. private communication) 
suggest that source intrinsic emission processes are responsible 
for the NIR spectral shape rather than extrinsic processes 
like extinction.
The fact that the dust source is most likely located behind SgrA* also 
suggests that is associated with the northern arm section of the mini-spiral
which is assumed to approach the central stellar cluster from behind
the plane of the sky in which SgrA* is located (Vollmer \& Duschl 2000).

Finally our investigation shows that most of the MIR flux density
seen towards the position of SgrA* is due to dust emission.
This suggests that the overall spectral shape of SgrA* is significantly 
less peaked in the FIR wavelength domain as suggested by the the upper limits. 
Combined with the results from our SSC modeling 
we find that one can expect that the intrinsic spectrum of SgrA* is
peaked at frequencies of a few THz.
The radio and submillimeter data show clear indications for 
variability. In general the flux density variations
are slow and occur on somewhat longer
timescales than the X-ray and IR variations.
The exact relation between the radio/sub-mm domain and the NIR/X-ray
domain still remains uncertain, due to the lack of sufficient 
simultaneous coverage. However, the amplitudes and time scales indicated
are consistent with a model in which the emitting 
material is expanding and cooling adiabatically.

Future observations will lead to improved statistics on the 
differences between simultaneous NIR and X-ray flares. Especially the
coupling to the mm-domain is of importance. Here, no simultaneous 
data are available so far.
Such observations will help to investigate 
whether individual mm-flare events are related to events in the
NIR or X-ray regime.
Upcoming simultaneous monitoring programs from the radio to the
X-ray regime will be required to further investigate the physical
processes that give rise to the observed IQ low NIR flux density state and flare phenomena
associated with SgrA* at the position of the massive black hole at the
center of the Milky Way.


\begin{acknowledgements}
This work was supported in part by the Deutsche Forschungsgemeinschaft
(DFG) via grant SFB 494.  \emph{Chandra} research is supported by NASA grants
NAS8-00128, NAS8-38252 and GO2-3115B.  We are grateful to all members
of the NAOS/CONICA and the ESO PARANAL team.

\end{acknowledgements}

\bibliography{gc}

\begin{thebibliography}{}


  \bibitem[Baganoff et al.(2001)]{Baganoff01} Baganoff, F.K., 
   Bautz, M.W., Brandt, W.N., et al. 2001, Nature, 413, 45

  \bibitem[Baganoff et al.(2003)]{Baganoff03} Baganoff, F. K., Maeda,
  Y., Morris, M., et al. 2003, ApJ 591, 891
  
  \bibitem[2003]{Baganoff} Baganoff, F. K., 2003,
  American Astronomical Society, HEAD meeting \#35, \#03.02

  \bibitem[2002]{Baganoff} Baganoff, F. K., et al., 2002, 201st AAS
  Meeting, \#31.08; Bulletin of the American Astronomical Society,
  Vol. 34, 1153

  \bibitem[2001]{ball} Ball, G.~H., Narayan, R. \& Quataert, E. 2001,
  ApJ, 552, 221

  \bibitem[1979]{BlandfordKonigl1979} Blandford \& K\"onigl 1979, \apj 232, 34-48

  \bibitem[1979]{BlandfordMcKee1979} 
  Blandford, R. D.; McKee, C. F., 1979, MNRAS 180, 343-371

  \bibitem[1985]{Bogdan1985} Bogdan, T.J., \& Schlickeiser, R., 1985, \aa, 143, 23

  \bibitem[2005]{Bower2005} Bower, G.C., Roberts, D.A., 
  Yusef-Zadeh, F., Backer, D.C., Cotton, W.D.,  Goss, W.M., Lang, C.C., Lithwick, Y.,
  2005, ``A Radio Transient 0.1 pc from Sagittarius A*,'' ApJ, in press

  \bibitem[2004]{BowerE04}
 {Bower}, G.~C., {Falcke}, H., {Herrnstein}, R.~M., {Zhao}, J., {Goss}, W.~M.,
  \& {Backer}, D.~C. 2004, Science, 304, 704

  \bibitem[2004] {BowerE05}
  {Bower}, G.~C., {Falcke}, H., {Wright}, M.~C., \& {Backer}, D.~C. 2005, \apjl,
  618, L29

  \bibitem[2003]{Bower2003}
  {Bower}, G.~C., {Wright}, M.~C.~H., {Falcke}, H., \& {Backer}, D.~C. 2003,
  \apj, 588, 331

  \bibitem[2002]{Bower2005} Bower, G.C., Falcke, H., Sault, R.J. and Backer, D.C., 2002, ApJ, 571, 843

  \bibitem[2005]{Bower2005} Bower, G.C.; Falcke, H., Wright, M.C., Backer, D.C., 2005, ApJ 618, L29

  \bibitem[1999]{blandford} Blandford, R., \& Begelman, M., 1999, MNRAS, 303, L1

  \bibitem[2000]{AA10000} Brandner, W., Rousset, G. Lenzen, et al., The Messenger 107, 1, 2002

  \bibitem[1985]{Bregman1985} Bregman, J.N., 1985, \aj, 288, 32
  

  \bibitem[2000]{AA10000} Cotera, A.~S., Simpson, J.~P., Erickson, E.~F.,
        Colgan, S.~W.~J., Burton, M.~G., Allen, D.~A., ApJ 510, 747, 1999


  \bibitem[2000]{AA10000} Diolaiti, E. Bendinelli, O., Bonaccini, D., Close, L., 
  Currie, D., Parmeggiani, G., A\&AS 147, 335-346, 2000

  \bibitem[1996]{eckart96} Eckart, A. \& Genzel, R. 1996, Nature 383, 415

  \bibitem[2000]{AA10000} Eckart, A.; Ott, T.; Genzel, R., A\&A 352, L22, 1999

  \bibitem[2002]{eckart02} Eckart, A., Genzel, R., Ott, T. and Schoedel,
  R. 2002, MNRAS, 331, 917

  \bibitem[2000]{AA10000} Eckart, A.; Baganoff, F. K.; Morris, M.; Bautz, M. W.; Brandt, W. N.; 
  Garmire, G. P.; Genzel, R.; Ott, T.; Ricker, G. R.; Straubmeier, C.; 
  Viehmann, T.; Schödel, R.; Bower, G. C.; Goldston, J. E., 2004, A\&A 427, 1

  \bibitem[2003]{eckart03} Eckart, A., Moultaka, J., Viehmann, T.  et
   al., 2003: Monitoring Sagittarius~A* in the MIR with the VLT. In:
   Proceedings of the Galactic Center Workshop, Nov. 3-8, 2002,
   Hawaii, A. Cotera, T. Geballe, S. Markoff, H. Falcke (editors),
   Astron. Nachr. 324


  \bibitem[2000]{AA10000} Eckart, A., Baganoff, F.K., Morris, M., Bautz, M.W., Brandt, W.N.,
   Garmire, G.P., Genzel, R., Ott, T., Ricker, G.R., Straubmeier, C.,
   Viehmann, T., Sch\"odel, R., Bower, G.C., Goldston, J.E., 2005,
  ''First Simultaneous NIR/X-ray Flare Detection from SgrA*'',
  Proceedings of a Conf. on 'Growing Black Holes' held in Garching, Germany,
  20-25 June, 2004

  \bibitem[2000]{AA10000} Eisenhauer, F.; Schödel, R.; Genzel, R.; Ott, T.;
  Tecza, M.; Abuter, R.; Eckart, A.; Alexander, T., 2003, ApJ 597, L121

  \bibitem[2000]{AA10000} Eisenhauer, F., Genzel, R., Alexander, T., Abuter, R., Paumard, T., Ott, T., 
    Gilbert, A., Gillessen, S., Horrobin, M., Trippe, S., Bonnet, H., Dumas, C., 
    Hubin, N., Kaufer, A., Kissler-Patig, M., Monnet, G., Str\"obele, S., 
    Szeifer, T., Eckart, A., Sch\"odel, R., \& Zucker, S., submitted to ApJ 2005

  \bibitem[2000]{AA10000} Figer, D.~F. , Becklin, E.~E., McLean, I.~S.,
        Gilbert, A.~M., Graham, J.~R., Larkin, J.~E.,
        Levenson, N.~A., Teplitz, H.~I., Wilcox, M.~K.,
        Morris, M., ApJ 533, L49, 2000


  \bibitem[1997]{genzel} Genzel, T., Eckart, A., Ott, T. \& Eisenhauer,
   MNRAS 1997, 291, 219

  \bibitem[2000]{genzel00} Genzel, R., Pichon, C., Eckart, A., Gerhard,
   O.E., Ott, T. 2000, MNRAS 317, 348

  \bibitem[2000]{AA10000} Gezari, S., Ghez, A.~M., Becklin, E.~E., Larkin, J., 
        McLean, I.~S., Morris, M., ApJ 576, 790, 2002

  \bibitem[Genzel et al.(2003)]{Genzel03} Genzel, R., Schoedel, R., Ott, T., et
   al. 2003, Nature, 425, 934

  \bibitem[1998]{ghez98} Ghez, A., Klein, B.L., Morris, M. \& Becklin,
   E.E. 1998, ApJ, 509, 678

  \bibitem[2000]{ghez00} Ghez, A., Morris, M., Becklin, E.E., Tanner,
   A. \& Kremenek, T.  2000, Nature 407, 349

  \bibitem[2003a]{ghez03a} Ghez, A. M., Duch\'ene, G., Matthews, K., et
  al. 2003a, ApJ, 586, L127

  \bibitem[2003b]{ghez03b} Ghez, A.M., Salim, S., Hornstein, S.D., et
    al. 2003b, ApJ, submitted, astro-ph/0306130

  \bibitem[Ghez et al.(2004a)]{Ghez04} Ghez, A.M., Wright, S.A., Matthews, K., et
    al. 2004a, ApJ 601, 159

  \bibitem[2004]{Ghez2004b} Ghez, A.M., Hornstein, S.D., Bouchez, A., Le Mignant, D., Lu, J., 
   Matthews, K., Morris, M., Wizinowich, P., Becklin, E.E., 2004b, AAS 205, 2406

  \bibitem[Ghez et al.(2005)]{Ghez05} Ghez, A.M., Salim, S., Hornstein, S. D., 
   Tanner, A., Lu, J. R., Morris, M., Becklin, E. E., Duch\^ene, G.,2005, ApJ 620, 744	

  \bibitem[Goldwurm et al.(2003)]{Goldwurm03} Goldwurm, A., Brion, E.,
  Goldoni, P. et al. 2003, ApJ, 584, 751

  \bibitem[1979]{gould} Gould, R.J., 1979, A\&A 76, 306

  \bibitem[2005]{Bower2005} Herrnstein, R.M., Zhao, J.-H., Bower, G.C., \& Goss, W.M., 2004, AJ, 127, 3399

  \bibitem[2000]{AA10000} Hornstein, S.~D., Ghez, A.~M., Tanner, A., Morris, M.,
    Becklin, E.~E., Wizinowich, P., ApJ 577, L9-L13, 2002


  \bibitem[{{Ho} {et~al.}(2004){Ho}, {Moran}, \& {Lo}}]{HoMoranLo04}
  {Ho}, P.~T.~P., {Moran}, J.~M., \& {Lo}, K.~Y. 2004, \apjl, 616, L1

  \bibitem[H{\o}g et al.(2000)]{hog} H{\o}g, E., Fabricius, C., Makarov, V.V. et al. 2000, A\&A, 355, L27

  \bibitem[2002]{igumenshchev} Igumenshchev, I.V., 2002, ApJ 577, 31

  \bibitem[2000]{Lagage} Lagage, P.-O., The final design of VISIR, the mid-infrared imager and 
   spectrometer for the VLT, SPIE Vol. 4008, pp. 1120 - 1131, March 2000

  \bibitem[2000]{AA10000} Lenzen, R., Hofmann, R., Bizenberger, P., Tusche, A.,
  Proc. SPIE Vol. 3354, p. 606-614, 
  Infrared Astronomical Instrumentation, Albert M. Fowler; Ed., 1998
  
  \bibitem[2000]{AA10000} Liu, S., Petrosian, V., Mason, G.M., 2006, submitted to ApJ (astro-ph/0506151).

  \bibitem[2000]{AA10000} Lutz, D., Feuchtgruber, H., Genzel, R., Kunze, D.,
        Rigopoulou, D., Spoon, H.~W.~W., Wright, C.~M.,
        Egami, E., Katterloher, R., Sturm, E., Wieprecht, E.,
        Sternberg, A., Moorwood, A.~F.~M., de Graauw, T.,
    A\&A 315, L269, 1996

  \bibitem[2001]{markoff} Markoff, S., Falcke, H., Yuan, F. \& Biermann, P.L.
   2001, A\&A, 379, L13

  \bibitem[1983]{marscher} Marscher, A.P. 1983, ApJ, 264, 296

  \bibitem[2005]{mau10000} Mauerhan, J.C.; Morris, M.; Walter, F.; 
  Baganoff, F.K.,  2005 ApJ 623, L25
  
  \bibitem[2001a]{melia01a} Melia, F. \& Falcke, H. 2001a, ARA\&A 39, 309

  \bibitem[2000]{AA10000} Moneti, A.; Stolovy, S.; Blommaert, J. A. D. L.;
     Figer, D. F.; Najarro, F., A\&A 366, 106, 2001


  \bibitem[1995]{narayan95} Narayan, R., Yi, I., \& Mahadevan, R. 1995,
  Nature, 374, 623

  \bibitem[2002]{narayan02} Narayan, R., Quataert, E., Igumenshchev, I.V., \&
    Abramowicz, M.A. 2002, ApJ, 577, 295

  \bibitem[Porquet et al.(2003)]{Porquet03} Porquet, D., Predehl, P., Aschenbach, et al.
    2003, A\&A 407, L17

  \bibitem[2000]{quataert00} Quataert, E., \& Gruzinov, A. 2000, ApJ 539, 809

  \bibitem[2003]{quataert03} Quataert, E., Astron. Nachr., Vol. 324,
  No. S1 (2003), Special Supplement "The central 300 parsecs of the
  Milky Way", Eds. A.  Cotera, H. Falcke, T. R. Geballe, S. Markoff,
  p. 435 (astro-ph/0304099)

  \bibitem[2000]{AA10000} Reid, Mark J., 1993, ARA\&A 31, 345

  \bibitem[Reid et al.(1999)]{Reid99} Reid, M.~J., Readhead, A.~C.~S., Vermeulen, R.~C., \&
  Treuhaft, R.~N. 1999, ApJ, 524, 816

  \bibitem[1998]{Rio} Rio Y. et al. , VISIR: The mid infrared imager and spectrometer for the VLT, 
    SPIE Vol. 3354, pp. 615-626, Kona, Hawaii, March 1998

  \bibitem[2000]{AA10000} Rousset, G., et al.,
  Adaptive Optical System Technologies II., 
  Edited by Wizinowich, Peter L.; Bonaccini, Domenico.  
  Proceedings of the SPIE, Volume 4839, pp. 140-149 (2003)


  \bibitem[2002]{schoedel02} Sch\"odel, R., Ott, T., Genzel, R. et
  al. 2002, Nature, 419, 694

  \bibitem[2003]{schoedel03} Sch\"odel, R., Genzel, R., Ott, et al. 2003,
  ApJ, 596, 1015

  \bibitem[2000]{AA10000} Scoville, N.~Z., Stolovy, S.~R.,  Rieke, M., Christopher, M.,
   Yusef-Zadeh, F., ApJ 594, 294, 2003

  \bibitem[2000]{AA10000} Serabyn, E.; Carlstrom, J.; Lay, O.; Lis, D. C.;
     Hunter, T. R.; Lacy, J. H.,  ApJ 490, L77, 1997

  \bibitem[2000]{AA10000} Stacey, G.~J., Nikola, T., Bradford, C.~M., 
  Hall, L., Bolatto, A.~D., Jackson, J.~M., Savage, M.~L., 
  Davidson, J.~A.,
  'SPIFI Imaging of the Galactic Center', in
  'The Dense Interstellar Medium in Galaxies', p.273, 2004

  \bibitem[2000]{AA10000} Stolovy, S.R.; Hayward, T. L.; Herter, T., ApJ 470, L45, 1996

  \bibitem[2000]{AA10000} Tanner, A., Ghez, A.~M., Morris, M., Becklin, E.~E., Cotera, A., 
  Ressler, M., Werner, M., Wizinowich, P., APJ 575, 860-870, 2002

  \bibitem[2000]{AA10000} Telesco, C. M.; Davidson, J. A.; Werner, M. W.,  ApJ 456, 541, 1996

  \bibitem[1966]{vdLaan} van der Laan, H., 1966, Nature 211, 1131

  \bibitem[2000]{AA10000} Vollmer, B. \& Duschl, W.~J., New Astronomy, 4, 581, 2000

  \bibitem[2000]{AA10000} Viehmann, T.; Eckart, A.; Moultaka, J.; Straubmeier, C.
  The Dense Interstellar Medium in Galaxies, Proceedings of the 
  4th Cologne-Bonn-Zermatt Symposium, Zermatt, Switzerland, 
  22-26 September 2003. Edited by S.Pfalzner, C. Kramer, C.
  Staubmeier, and A. Heithausen. Springer proceedings in physics,
  Vol. 91. Berlin, Heidelberg: Springer, 2004, p.303

  \bibitem[2000]{AA10000} Viehmann, T.; Eckart, A.; Schödel, R.; Moultaka, J.;
      Straubmeier, C.; Pott, J.-U., A\&A 433, 117, 2005

  \bibitem[2002]{weisskopf} Weisskopf, M.~C., Brinkman, B., Canizares,
  C., et al. 2002, PASP, 114, 1

  \bibitem[2002]{yuan02} Yuan, F., Markoff, S. \& Falcke, H. 2002,
  A\&A, 854, 854

  \bibitem[2003]{yuan03} Yuan, F., Quataert, E.  \& Narayan, R. 2003,
  ApJ, 598, 301

  \bibitem[2004]{yuan04} Yuan, F., Quataert, E. \& Narayan, R. 2004,
  ApJ,  606, 894

  \bibitem[2003]{Zhao2003} {Zhao}, J., {Young}, K.~H., {Herrnstein}, R.~M., {Ho}, P.~T.~P., {Tsutsumi},
  T., {Lo}, K.~Y., {Goss}, W.~M., \& {Bower}, G.~C. 2003, \apjl, 586, L29

  \bibitem[2005]{zhao2003} Zhao, J.-H., Young, K.H.,  Herrnstein, R.M., Ho, P.T.P., Tsutsumi, T., Lo,
  K.Y., Goss, W.M. \& Bower, G.C., 2003, ApJL, 586, L29.

  \bibitem[2005]{zhao2004} Zhao, J.-H., Herrnstein, R.M., Bower, G.C., Goss, W.M., \& Liu, S.M., 2004, ApJL, 603, L85.


\end{thebibliography}

\newpage

%

\begin{table}
\centering
{\begin{small}
\begin{tabular}{ccccc}
\hline
Telescope & Instrument & Energy/$\lambda$ & UT Start Time & UT Stop Time \\
Observing ID & & & &  \\
\hline
\emph{Chandra} & ACIS-I    & 2-8 keV & 05 JUL 2004 22:38:26  & 06 JUL 2004 12:56:59 \\
\emph{Chandra} & ACIS-I    & 2-8 keV & 06 JUL 2004 22:35:12  & 07 JUL 2004 12:53:45 \\
1 VLT UT~4       & NACO    & 1.7~$\mu$m  & 06 JUL 2004 02:47:11 & 06 JUL 2004 03:26:48 \\      
2 VLT UT~4       & NACO    & 2.2~$\mu$m  & 06 JUL 2004 03:48:53 & 06 JUL 2004 07:05:59 \\      
3 VLT UT~4       & NACO    & 3.8~$\mu$m  & 06 JUL 2004 07:17:10 & 06 JUL 2004 08:42:52 \\      
4 VLT UT~4       & NACO    & 2.2~$\mu$m  & 06 JUL 2004 23:19:39 & 07 JUL 2004 04:16:37 \\      
5 VLT UT~4       & NACO    & 3.8~$\mu$m  & 07 JUL 2004 23:54:48 & 08 JUL 2004 00:45:44 \\      
6 VLT UT~4       & NACO    & 2.2~$\mu$m  & 08 JUL 2004 00:53:32 & 08 JUL 2004 06:53:46 \\      
SMA-A            & 340 Pol & 890~$\mu$m  & 05 JUL 2004 06:59:30 & 05 JUL 2004 12:24:18 \\
SMA-B            & 340 Pol & 890~$\mu$m  & 06 JUL 2004 06:03:17 & 06 JUL 2004 12:26:48 \\
SMA-C            & 340 Pol & 890~$\mu$m  & 07 JUL 2004 05:47:12 & 07 JUL 2004 12:28:58 \\
VLA-A            & ...     & 0.7~cm      & 06 Jul 2004 04:41:24 & 06 Jul 2004 09:09:06 \\
VLA-B            & ...     & 0.7~cm      & 07 Jul 2004 04:37:46 & 07 Jul 2004 09:02:26 \\
VLA-C            & ...     & 0.7~cm      & 08 Jul 2004 04:39:16 & 08 Jul 2004 08:58:27 \\
\hline
\end{tabular}
\end{small}}
\caption{Observation Log.}
\label{log}
\end{table}

\newpage
\FloatBarrier
\begin{table}
\centering
{\begin{small}
\begin{tabular}{ccccccc}
\hline
Observing & $\lambda$ & DIT & NDIT & N & Pixel Scale & Seeing  \\
ID &  & &  & & &  \\
\hline
1  &  1.7\,$\mu$m & 15\,s  & 2   & 50  & 0.013$''$     & $\sim$$1.1''$  \\
2  &  2.2\,$\mu$m & 15\,s  & 2   & 217 & 0.027$''$     & $\sim$$1.0-1.5''$ \\
3  &  3.8\,$\mu$m & 0.2\,s & 150 & 100 & 0.027$''$     & $\sim$$1.3-1.8''$  \\
4  &  2.2\,$\mu$m & 30\,s  & 1   & 330 & 0.027$''$     & $\sim$$1.0-1.8''$  \\
5  &  3.8\,$\mu$m & 0.2\,s & 150 & 60  & 0.027$''$     & $\sim$$0.7''$  \\
6  &  2.2\,$\mu$m & 30\,s  & 1   & 391 & 0.027$''$     & $\sim$$0.7''$  \\
\hline
\end{tabular}
\end{small}}
\caption{Details of NIR observations. ``Observing ID'' refers to the number of
the data set in column one of Table~\ref{log}. $\lambda$ is the
central wavelength of the broad-band filter used. DIT is the detector
integration time in seconds. NDIT is the number of exposures of
integration time DIT that were averaged on-line by the instrument. N
is the number of images taken. The total integration time amounts to
DIT$\times$NDIT$\times$N. Seeing is the value measured by the
Differential Image Motion Monitor (DIMM) on Paranal at visible
wavelengths. It provides a rough impression on atmospheric conditions
during the observations.
\label{NIRObs}}
\end{table}


\newpage
\FloatBarrier
\begin{table}
\centering
{\begin{small}
\begin{tabular}{ccccc}
\hline
Observing & $\lambda$ & RA [$''$] & DEC [``] & Flux [mJy] \\
ID & & & &  \\
\hline
1 & 1.7\,$\mu$m & 0.004$\pm$0.004 & 0.009$\pm$0.005 & $\le$3.1 \\ 
2 & 2.2\,$\mu$m & 0.011$\pm$0.007 & 0.012$\pm$0.013 & $\le$3.9 \\ 
3 & 3.8\,$\mu$m & 0.017$\pm$0.009 & -0.017$\pm$0.004 &$\le$34  \\ 
4 & 2.2\,$\mu$m & 0.004$\pm$0.003 & 0.005$\pm$0.004 & 5.2$\pm$1.8 \\ 
5 & 3.8\,$\mu$m & -0.022$\pm$0.004 & -0.009$\pm$0.005 & 17$\pm$4 \\ 
6 & 2.2\,$\mu$m  & 0.003$\pm$0.004 & 0.002$\pm$0.003 & 2.8$\pm$1.1 \\ 
\hline
\end{tabular}
\end{small}}
\caption{Average position and flux of Sgr~A* as obtained from the NIR data.
The first column lists the observation ID (see Table~\ref{log}), second
and third columns the position of Sgr~A* (average of all exposures)
relative to its nominal position (Eisenhauer et al. 2003), and the
fourth column the measured overall flux and standard deviation of Sgr~A*
during the particular observing session. The
data were obtained by aperture measurements with a $\sim$50\,mas
radius circular aperture on the nominal position of Sgr~A*. The flux
measurements were corrected for extinction (see text). For data set~2,
the average values are extracted from the first 120 exposures because
the exposures obtained later were of very low quality.
\label{SgrAdata}}
\end{table}


\newpage
\FloatBarrier
\begin{table}[!h]
\begin{center}
{\begin{small}
\begin{tabular}{ccccccccc} \hline
X-ray  &Start & Stop & FWZP & FWHM & Extraction & Total & Flare & IQ state \\
flare ID      &      & (min)& (min)& & Radius  & &  \\
& & & &  &(arcsec) &  & & \\ \hline
$\phi$1& 06 JUL 02:22:00 & 06 JUL 03:12:00 & 50$\pm$10 & -         & 1.5 & 14 $\pm$ 7 &  8 $\pm$ 7 & 5.9 $\pm$ 1.0  \\
$\phi$2& 06 JUL 23:09:00 & 06 JUL 23:34:00 & 25$\pm$10 & -         & 1.5 & 18 $\pm$ 7 & 11 $\pm$ 7 & 7.0 $\pm$ 1.0  \\
$\phi$3& 07 JUL 02:57:24 & 07 JUL 03:40:00 & 42$\pm$5  & 10$\pm$5  & 1.5 & 86 $\pm$ 7 & 79 $\pm$ 7 & 7.0 $\pm$ 1.0  \\
$\phi$4& 07 JUL 03:40:00 & 07 JUL 04:00:00 & 20$\pm$5  & -         & 1.5 & 20 $\pm$ 7 & 13 $\pm$ 7 & 7.0 $\pm$ 1.0  \\
\end{tabular}
\end{small}}
\end{center}
\caption{X-ray Flare Count Rates:
Given are the peak times and peak 
ACIS-I count rates in $\times 10^{-3}$ cts s$^{-1}$ in 2--8 keV band of the total flare 
emission and the flare emission corrected for the count rate during
the IQ state. We also list the estimated start and stop times as well as the full width at zero power (FWZP) and
full width at half maximum (FHWM) values, as well as the peak and IQ flux densities.
The candidate X-ray flare events $\phi$2 and $\phi$4 coincide with
significant NIR flares (labeled I and IV in Tab.~\ref{NIRobs}).
The candidate  X-ray flux density increase $\phi$1 is similar to $\phi$2.
For the weak candidate flare events $\phi$1, $\phi$2, and $\phi$4 we only give estimates of FWZP.
\label{flareprop}}
\end{table}


\newpage
\FloatBarrier
\begin{table}[!h]
\begin{center}
{\begin{small}
\begin{tabular}{ccccccccc} \hline
MIR&  date   &    time  &    flux  & seeing   \\
ID &         &          &   density  &FWHM   \\
   &         &          &   (mJy)    &(pixel)   \\
 \hline
001& 2004-05-08& 09:07:47&   25.2&    0.27   \\
002& 2004-05-08& 09:09:52&   23.2&    0.27   \\
003& 2004-05-08& 09:10:27&   24.2&    0.27   \\
004& 2004-05-08& 09:12:35&   23.3&    0.25   \\
005& 2004-05-08& 09:13:10&   24.8&    0.25   \\
006& 2004-05-08& 09:21:12&   21.6&    0.27   \\
007& 2004-05-08& 09:23:15&   24.5&    0.28   \\
008& 2004-05-08& 09:23:51&   24.2&    0.23   \\
009& 2004-05-08& 09:25:56&   24.5&    0.25   \\
010& 2004-05-08& 09:46:54&   25.6&    0.24   \\
011& 2004-05-08& 09:51:53&   31.4&    0.28   \\
012& 2004-05-08& 09:53:57&   32.0&    0.29   \\
013& 2004-05-08& 09:54:34&   20.6&    0.27   \\
014& 2004-05-08& 09:56:40&   32.3&    0.30   \\
015& 2004-05-08& 09:57:15&   27.3&    0.27   \\
016& 2004-05-09& 08:16:59&   30.8&    0.33   \\
017& 2004-05-09& 08:19:08&   47.4&    0.43   \\
018& 2004-05-09& 08:19:46&   46.4&    0.37   \\
019& 2004-05-09& 08:21:52&   56.9&    0.38   \\
020& 2004-05-09& 08:22:26&   56.7&    0.38   \\
021& 2004-05-10& 05:59:04&   23.7&    0.38   \\
\end{tabular}
\end{small}}
\end{center}
\caption{8.6$\mu$m flux densities:
Flux densities are given in mJy.
FWHM is given in acrseconds.
The flux densities are dereddened using $A_{8.6} = 1.75$.
The calibration is described in the text.
\label{VISIRfluxes}}
\end{table}

\newpage
\FloatBarrier
\begin{table}
\centering
{\begin{tiny}
\begin{tabular}{cccccccccc}
\hline
Observing & IR Flare & $\lambda$    & Start              & Stop            & FWZP  & FWHM & peak flux     & IQ flux \\
ID  & ID   &              &                    &                 & (min) & (min)& (mJy)         & (mJy) \\
\hline
1  &     & 1.7\,$\mu$m &  &  &  &  & $<$4.1 & 2.60 $\pm$ 0.5 \\
2  &     & 2.2\,$\mu$m &  &  &  &  &  & 3.20 $\pm$ 0.7 \\
3  &     & 3.8\,$\mu$m &  &  &  &  &  & 23.0 $\pm$ 11.0 \\
4  & I   & 2.2\,$\mu$m &$<$ 06 JUL 23:19:39 & 06 JUL 23:32:00 & $>$12 & $>$6 & $\ge$5.7      & 3.00 $\pm$ 1.0 \\
4  & II  & 2.2\,$\mu$m &    07 JUL 00:29:00 & 07 JUL 02:39:00 & 130   & $>$70& $\sim$ 3.0    & 3.00 $\pm$ 1.0 \\
4  & III & 2.2\,$\mu$m &    07 JUL 02:59:00 & 07 JUL 03:39:00 &  45   & 25-35& 6.0$\pm$ 1.5  & 3.00 $\pm$ 1.0 \\
4  & IV  & 2.2\,$\mu$m &    07 JUL 02:39:00 & 07 JUL 03:59:00 &  20   & $\sim$10& 5.0$\pm$ 1.5  & 3.00 $\pm$ 1.0 \\
5  &     & 3.8\,$\mu$m &  &  &  &  &  & 17.0 $\pm$ 4.0 \\
6  & V   & 2.2\,$\mu$m &    08 JUL 02:33:00 & 08 JUL 04:03:00 &  95   & 40-55& 4.0$\pm$1.0   & 3.00 $\pm$ 1.0 \\
\hline
\end{tabular}
\end{tiny}}
\caption{Emission properties of the NIR flare events from the infrared counterpart of SgrA*.
For each observing session we give the estimated IQ flux, which 
corresponds to the mean flux during its low flux density state
during that session. 
The peak flux densities corrected for the IQ state flux. 
The flux densities are dereddened 
using $A_{H} = 4.3$,  $A_{K} = 2.8$, and  $A_{L'} = 1.8$.
In case of a flare event detection we give the full zero start and stop times, the full zero width 
at the corresponding zero points (FWZP) and the full width at half maximum of the flare events. 
The time period listed with the infrared flare event II corresponds to a time of 
slightly increased source activity.
\label{NIRobs}}
\end{table}

\newpage
\FloatBarrier
\begin{table}[!h]
\begin{center}
{\begin{small}
\begin{tabular}{ccccclccc} \hline
X-ray   & NIR & X-ray flux    &NIR flux  & NIR  & spectral\\
 flare  & flare & density     & density    & band & index \\
 ID     &  ID   & (nJy)       & (mJy)      &      & $\alpha_{NIR/X-ray}$ \\
\hline
$\phi$1 &     & 22$\pm$27   & 2.6$\pm$0.5&  H   & -1.34$\pm$0.2 \\
$\phi$2 &  I  & 31$\pm$27   & $\ge$5.7   &  K   & -1.35$\pm$0.2 \\
        & II  & $<$20       & $\sim$3.0  &  K   & $<$-1.34 \\
$\phi$3 &III  & 223$\pm$27  & 6$\pm$1.5  &  K   & -1.12$\pm$0.05 \\
$\phi$4 & IV  &  37$\pm$27  & 5$\pm$1.5  &  K   & -1.35$\pm$0.2 \\
        & V   & -           & 4$\pm$1.0  &  K   &  - \\
\end{tabular}
\end{small}}
\end{center}
\caption{NIR/X-ray Flare Flux Densities.
Given are the peak flux densities of the flares detected in the individual 
wavelength bands. The spectral index is calculated assuming band centers
of 2.2$\mu$m and 1.6$\mu$m in the near-infrared and 4~keV in the X-ray domain.
The X-ray flares $\phi$2, $\phi$3, and $\phi$4 have been detected simultaneously in 
the NIR. For flare $\phi$1 only an upper limit in the H-band is available.
For flare V no X-ray data exist.
See comments to the candidate X-ray flare events $\phi$1, $\phi$2, and $\phi$4
in the text and in in Tab.~\ref{flareprop}.
\label{flares}}
\end{table}


\newpage
\FloatBarrier
\begin{table}
\centering
{\begin{tiny}
\begin{tabular}{cccccccccc}
\hline
Observing & IR Flare & $\lambda$    & Start              & Stop            & FWZP  & FWHM & peak flux     \\
ID  & ID   &              &                    &                 & (min) & (min)& (Jy)          \\
\hline
SMA-A  & SMA1& 890\,$\mu$m &    05 JUL 07:00 &      05 JUL 10:30 & 210   & 120   &       3.5       \\
SMA-A  & SMA2& 890\,$\mu$m &    05 JUL 11:00 & $\ge$05 JUL 12:10 & -     & -     &  $\ge$3.2       \\
SMA-B  & SMA3& 890\,$\mu$m &    06 JUL 07:45 &      06 JUL 10:00 & 135   & $<$60 &       3.4       \\
SMA-C  & SMA4& 890\,$\mu$m &$<$ 07 JUL 06:00 &      07 JUL 08:00 &  -    & -     &  $\ge$3.5       \\
SMA-C  & SMA5& 890\,$\mu$m &    07 JUL 08:00 & $\ge$07 JUL 12:00 &  -    & -     &       3.3       \\
VLA-B  & VLA1& 7~mm        &$<$ 07 JUL 04:40 & $>$  07 JUL 08:50 &       &       &    $>$0.5       \\
\hline
\end{tabular}
\end{tiny}}
\caption{Emission properties of the radio flare events from SgrA*.
In case of a flare event detection we 
we give the estimated peak flux density,
the full zero start and stop times, the full zero width 
at the corresponding zero points (FWZP) and the full width at half maximum of the flare events. 
For the VLA we give the excess flux density over the mean flux density 
measured on 06 and 08 July.
\label{RADIOobs}}
\end{table}


\FloatBarrier
\begin{table}
\centering
{\begin{small}
\begin{tabular}{ccccccccccc}
\hline
model & S$_{NIR}$ & S$_{NIR}$ &  S$_{X-ray}$& B     & $\nu$$_{max,obs}$ & S$_{max,obs}$ & size     & $\alpha$$_{NIR/X-ray}$  &c/$\nu$$_{0}$ \\
ID & synchr.   &  SSC      &             &       &                   &               &          &                &        \\
  & (mJy)     &  (mJy)    &  (nJy)      & (G)   & (GHz)             &  (Jy)         &($\mu$as) &                 & ($\mu$m)       \\ \hline
IQ1      &
0.8   &  2.0  & $<$18   & 17 & 820          & 3.9          & 7.9     & 1.3 & -\\
IQ2   &
-     &  3.0  & $<$27   & 5  & 820          & 7.7          & 8.1     & 1.3 & -\\
IQ3   &
(25)1 &  0.1 & $<$17   & 80 & 1000         & 16.0         & 17.5     & 1.0  & 15\\
IQ4   &
2.0   &  0.02 & $<$15   & 38 & 1500         & 0.33         & 1.5     & 0.8 & -\\
F1    &
(18) 6&  1.13 & 65      & 68 & 1640         & 11.5         & 8.5     & 1.1 & 5\\
F2      &
(300) 6&  0.1  & 230    & 100 & 1640         & 5.7         & 8.5     & 0.4 & 90 \\
\hline
\end{tabular}
\end{small}}
\caption{Parameters for representative models in agreement with the IQ (IQ1-IQ4) and 
flare states (F1 and F2) observed towards SgrA*.
These models are plotted in Figs.~\ref{Fig:1NIRMIR} and  \ref{Fig:2NIRMIR}.
Listed are the NIR flux density contributions from the synchrotron and SSC 
part of the spectrum 
as well as the SSC X-ray flux density 
(columns 2 to 4). In the following columns we list  the magnetic field 
strength $B$, 
observed cutoff frequency $\nu_{m, obs}$ and flux density $S_{m, obs}$ 
of the synchrotron spectrum, 
size $\theta$ of the source component and the spectral index $\alpha$ 
of the synchrotron component.
All models except IQ2 provide synchrotron emission for frequencies up to
$\nu_2$$\sim$200THz. We assume that the cutoff in the energy spectrum
of the relativistic electrons can be represented via an exponential
cutoff in the observed synchrotron spectrum proportional to
$exp[-(\nu/\nu_0)^{0.5}]$ with the effective cutoff frequency $\nu_0$.
In the first column the NIR flux densities are given without (in brackets) and with
modulation by the exponential cutoff.
\label{tab:models}}
\end{table}


\FloatBarrier
\newpage


\FloatBarrier

\FloatBarrier

\FloatBarrier


\end{document}